\documentstyle[12pt]{article}
\def\ie{{\em i.e.}}
\def\eg{{\em e.g.}}

\def\beq{\begin{equation}}
\def\eeq{\end{equation}}

\def\Tr#1{{\rm Tr}\,#1}
\catcode`\@=11 
\def\coeff#1#2{{\textstyle{#1\over #2}}}

\def\VEV#1{\left\langle #1\right\rangle}
\def\vev#1{\langle #1\rangle}
\def\lsim{\mathrel{\mathpalette\@versim<}}
\def\gsim{\mathrel{\mathpalette\@versim>}}
\def\@versim#1#2{\vcenter{\offinterlineskip
    \ialign{$\m@th#1\hfil##\hfil$\crcr#2\crcr\sim\crcr } }}
\def\etal{{\em et. al.}}
\def\JL{J. L. Lopez}
\def\DVN{D. V. Nanopoulos}
\def\AZ{A. Zichichi}

\def\r#1{$\bf#1$}
\def\rb#1{$\bf\overline{#1}$}

\def\t1{{\tilde 1}}

\def\eV{\,{\rm eV}}

\def\GeV{\,{\rm GeV}}
\def\TeV{\,{\rm TeV}}
\def\y{\,{\rm y}}
\def\cm{\,{\rm cm}}

\def\wt{\widetilde}
\def\lra{\leftrightarrow}
\def\to{\rightarrow}

\def\ipb{\,{\rm pb}^{-1}}
\def\ifb{\,{\rm fb}^{-1}}

\def\NPB#1#2#3{Nucl. Phys. B {\bf#1} (19#2) #3}
\def\PLB#1#2#3{Phys. Lett. B {\bf#1} (19#2) #3}
\def\PRD#1#2#3{Phys. Rev. D {\bf#1} (19#2) #3}
\def\PRL#1#2#3{Phys. Rev. Lett. {\bf#1} (19#2) #3}
\def\PRT#1#2#3{Phys. Rep. {\bf#1} (19#2) #3}
\def\MODA#1#2#3{Mod. Phys. Lett. A {\bf#1} (19#2) #3}
\def\IJMP#1#2#3{Int. J. Mod. Phys. A {\bf#1} (19#2) #3}

\def\hepph#1{{\tt hep-ph/#1}}
\def\hepth#1{{\tt hep-th/#1}}
\def\hepex#1{{\tt hep-ex/#1}}
\def\astroph#1{{\tt astro-ph/#1}}

\begin{document}
\begin{flushright}
\baselineskip=12pt
DOE/ER/40717--23\\
{\tt hep-ph/9601208}\\
\end{flushright}
\vspace{1cm}
\begin{flushleft}
\Huge\bf Supersymmetry:
\end{flushleft}
\begin{flushright}
\Large\bf From the Fermi Scale to the Planck Scale\footnote{To appear in
Reports on Progress in Physics.}\\
\end{flushright}
\vglue 2cm
\begin{center}
{\Large\bf Jorge L. Lopez\\}
\vglue 1cm
{\large
Department of Physics\\
Bonner Nuclear Lab\\
Rice University\\
6100 Main Street\\
Houston, TX 77005\\}
\baselineskip=12pt
\end{center}

\vglue 1cm
\begin{abstract}
The physics of supersymmetry is reviewed from the perspective of physics at
ever increasing energies. Starting from the minimal supersymmetric extension of
the Standard Model at the electroweak scale, we proceed to higher energies
seeking to understand the origin of the many model parameters. Supersymmetric
grand unification, supergravity, and superstrings are introduced sequentially,
and their contribution to the sought explanations is discussed. Typical
low-energy supersymmetric models are also presented, along with their possible
experimental consequences via direct and indirect processes at high-energy
physics experimental facilities.
\end{abstract}

\vspace{1cm}
\begin{flushleft}
\baselineskip=12pt
DOE/ER/40717--23\\
January 1996
\end{flushleft}
\newpage

\tableofcontents
\newpage

\setcounter{page}{1}
\pagestyle{plain}
\baselineskip=14pt

\section{Introduction}
\label{sec:Intro}
\subsection{The Standard Model and beyond}
With the commissioning of Fermilab's Tevatron proton-antiproton collider in
1988 and CERN's LEP electron-positron collider in 1989, the Standard Model of
the strong and electroweak interactions has received overwhelming experimental
support, with some predictions checked to accuracies as high as 0.1\%. Among
the several experimental measurements, perhaps the most impressive ones have
been the precise determination of the masses of the $W$ ($M_W=80.41\pm0.18\GeV$
\cite{CDFW}) and $Z$ ($M_Z=91.1884\pm0.0022\GeV$
\cite{LEP}) gauge bosons which mediate the electroweak interactions, and the
related weak mixing angle ($\sin^2\theta_W=0.23186\pm0.00034$), as well as the
number of light neutrino species ($N_\nu=2.991\pm0.016$). This body of data
agrees with the predictions of the Standard Model in basically every instance,
and helps ``clean up" the mass range below ${1\over2}M_Z$, where no new physics
has been observed. Such degree of purity of the Standard Model below the
electroweak scale needs to be naturally accommodated by any of its proposed
extensions.

The most recent confirmation of the Standard Model has been the discovery at
the Tevatron (1995) of the long sought-for top quark by the CDF and D0
Collaborations \cite{Top}. The mass of the top quark, which early on had been
theorized to be as low as $20\GeV$, has experimentally turned out to be some
ten times larger. The Standard Model, in fact, does not predict the mass of the
top quark, or the mass of any other quark or lepton, or the quark mixing
angles. Similarly there is no explanation for the observed number of fermion
families (three), the quantization of the electric charge, the magnitude of the
weak mixing angle, the dynamical origin of the electroweak symmetry breaking,
the mass of the Higgs boson, etc. The Standard Model also lacks some appealing
features, such as neutrino masses, unified strong and electroweak symmetry and
gravity, matter instability (proton decay), and a cold dark matter candidate.
Finally, the Standard Model has some subtle problems when extrapolated to
very high energies: the electromagnetic (QED) and top-quark Yukawa couplings
encounter a Landau pole (\ie, they become very large) at sufficiently high
energies. These various shortcomings of the Standard Model require theoretical
explanation, although they do not detract from the fact that the Standard Model
is an excellent effective theory for energies $\lsim{\cal O}(100\GeV)$.

In this review we will argue that the Standard Model parameters and features
are most clearly understood when the energy scale of the interactions is
extrapolated to higher values ($Q\gg M_Z$). This extrapolation is postulated to
uncover larger symmetries which correlate various model parameters. These
larger theories are (much) more constrained than the Standard Model, and
therefore have (much) greater predictive power.

The basic underlying {\em assumptions} that we make in considering different
scales is that physical quantities at different mass scales (\eg, the
electroweak scale $Q\sim M_Z\sim10^2\GeV$ and the Planck scale $Q\sim
M_{Pl}\sim10^{19}\GeV$) are connected in a {\em calculable} and {\em natural}
way. In the realm of quantum field theory, specific relations between physical
quantities at different mass scales are required, as dictated by the {\em
renormalization group} invariance of the theory. These renormalization group
equations (RGEs) depend on the nature of the theory, and can be derived
explicitly in all cases of interest. Starting from a weakly interacting
theory at low mass scales (the Standard Model), these equations can be used to
evolve the model parameters to larger mass scales. Our calculability assumption
 implies that the model should remain perturbative (\ie, weakly interacting)
all the way up to the highest mass scales to be considered. The RGEs themselves
do not guarantee this, as for example in the case of the QED gauge coupling
(the
fine-structure constant $\alpha$), which becomes large at very high mass
scales. The calculability assumption is satisfied in this case by the existence
of the Planck scale, which effectively cuts off the growth of
$\alpha$.

Our naturalness assumption is also not guaranteed by the RGEs. Very heavy
particles can creep into the low-energy world through their appearance in
self-energy loop diagrams of fundamental scalar particles (like the Higgs
boson). These diagrams have a quadratic dependence on the high-energy cutoff (a
quadratic divergence), and once renormalized yield a finite contribution to the
scalar mass shift proportional to, \eg, $(M_{Pl}/M_Z)^2\sim10^{34}$. These
extremely large mass shifts can be compensated by an equally extreme and
unnatural fine-tuning of the renormalized model parameters. This is the {\em
gauge hierarchy problem}, which pervades any attempt at extrapolation to very
high mass scales of theories with fundamental scalar particles, and violates
our naturalness assumption.
Two solutions to this problem have been proposed: either there are no
fundamental scalar particles, or the high-energy behavior of the scalar
self-energy diagrams is somehow alleviated. The first solution leads to the
ideas of technicolor and compositeness, which have as a principal drawback
their general lack of calculability, and therefore violate our first
assumption. (Nonetheless, models based on these ideas are regularly considered,
although with limited phenomenological success.)

The second solution to the gauge hierarchy problem is based on the idea
of {\em Supersymmetry}. In a theory with fundamental scalars, supersymmetry
tames the quadratic divergences by predicting the existence of a
``superpartner"
to each particle, with the same mass and gauge quantum numbers, but with spin
differing by half-a-unit. The new fermionic superpartners of scalars and gauge
bosons contribute to the Higgs-boson self-energy loop diagrams and, because of
their different spin-statistics but same mass and gauge quantum numbers, lead
to an automatic cancellation of the quadratic divergences. The scalar mass
shifts vanish altogether in the limit of exact supersymmetry. However,
supersymmetry cannot be an exact symmetry of Nature, since otherwise light
superpartners of the quarks and leptons would have been observed. The breaking
of supersymmetry manifests itself via mass splittings between superpartners.
Such mass splittings contribute to the Higgs-boson mass shifts, and should not
exceed $\sim{\cal O}(100\GeV)$ if the gauge hierarchy problem is not to be
reintroduced, otherwise our naturalness assumption would again be violated.

Further theoretical motivation for supersymmetry is found in the form of
supergravity, which provides an effective description of quantum gravity at
energies below the Planck scale. Also, spacetime (four-dimensional)
supersymmetry is a typical prediction of string theory, whereas world-sheet
(two-dimensional) supersymmetry is required in order for particles with
half-integer spin to exist. Phenomenologically, supersymmetry solves the gauge
hierarchy problem and gives meaning to grand unification, which agrees
well with the low-energy measurements of the Standard Model gauge couplings.
Supersymmetry also explains dynamically the breaking of the electroweak
symmetry via radiative corrections, and predicts the existence of a light Higgs
boson ($m_h\lsim150\GeV$). Supersymmetric models typically provide a good
candidate for cosmological (cold) dark matter: the lightest supersymmetric
particle (LSP) -- the neutralino. Experimentally, supersymmetric models predict
the existence of many ($\gsim30$) new elementary particles -- the superpartners
of the Standard Model particles -- which should be accessible at present and
near-future high-energy accelerators via distinct signatures, such as missing
energy and low-background multi-leptons signals.

\begin{figure}[p]
\vspace{7in}
\includegraphics{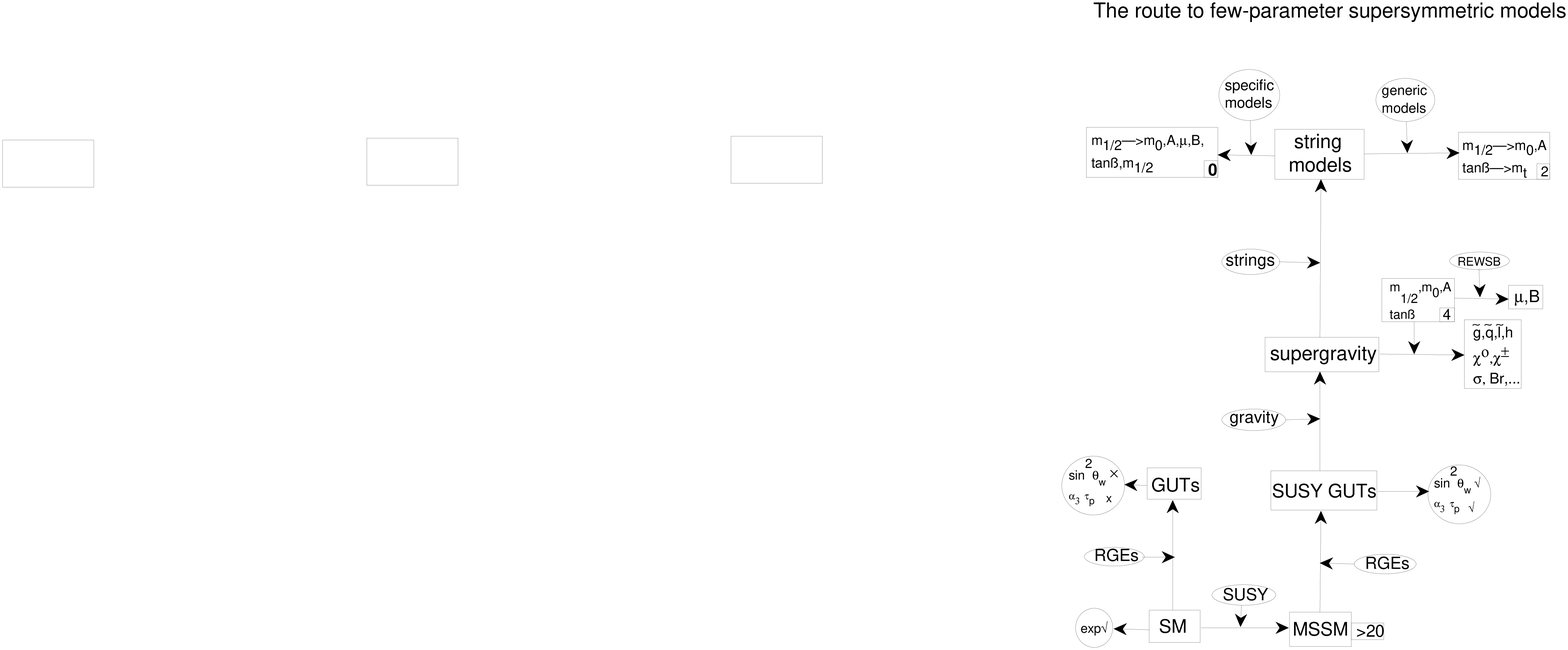}
\caption{Supersymmetry: From the Fermi Scale to the Planck Scale. The explicit
numbers (20,4,2,0) indicate the number of new parameters required to describe
the corresponding model.}
\label{fig:BigPicture}
\end{figure}

\subsection{Supersymmetry and unification}
Our first step away from the Standard Model entails the introduction of
low-energy supersymmetry,\footnote{The discussion that starts here and
continues through to the end of this section outlines the contents of this
review and the logic that underlies it, and is graphically summarized in
the form of a flow chart or road map in Fig.~\ref{fig:BigPicture}.} whereby
each particle in the Standard Model is accompanied by a superpartner. As such
this step does not appear to accomplish much in our quest for explanation of
the many Standard Model parameters. On the contrary, the number of parameters
is greatly increased by the addition of
more than 30 parameters needed to describe the new particle masses. Because of
the nature of supersymmetry one also needs to enlarge the Higgs sector from one
to two Higgs doublets. However, since supersymmetry ``commutes" with the gauge
interactions, no new unknown couplings are introduced. The resulting model is
referred to as the Minimal Supersymmetric Standard Model (MSSM). Low-energy
supersymmetry does provide a sometimes overlooked benefit in the Higgs sector.
In the Standard Model the mass of the Higgs boson derives from a quartic
coupling in the scalar potential, and is thus unconstrained, although
considerations of a weakly interacting Higgs sector suggest that
$M_H\lsim(500-1000)\GeV$ \cite{DJL}. In supersymmetry the quartic coupling is
not an independent parameter, but rather a simple function of the gauge
couplings, which are known and are not large. This prediction implies an upper
bound on the lightest Higgs boson in the MSSM of $m_h\lsim130\GeV$. That is,
the mass of the (lightest) Higgs boson derives from the electroweak
interactions, in contrast with the Standard Model, where a new sector of the
theory needs to be postulated.

To make real gains in explaining the Standard Model features, we extrapolate
the MSSM to higher energies. This procedure entails ``running" or scaling up
the model parameters (gauge and Yukawa couplings and mass parameters) by means
of the RGEs, hoping to uncover simple relations among the parameters at
sufficiently high mass scales. Since this scaling is logarithmic, significant
changes in the magnitude of the parameters requires order-of-magnitude changes
in the mass scales. Using the precise LEP measurements of the
Standard Model gauge couplings, it has been shown that this exercise yields a
convergence of the gauge couplings towards a single value at the scale
$Q=M_{\rm GUT}\sim10^{16}\GeV$. This phenomenon is {\em not} observed in
the Standard Model, which evolves differently because of its different particle
content. The convergence of the gauge couplings in the MSSM is
very suggestive of a larger symmetry being encountered at such large mass
scales: a {\em grand unified theory} (GUT). In fact, this result has been taken
by many as indirect evidence for the existence of (low-energy) supersymmetry.
However, the MSSM alone does not lead to true unification since, without the
new larger symmetry, the gauge couplings would  diverge when extrapolated
past $Q=M_{\rm GUT}$.

If a GUT is indeed present above $Q=M_{\rm GUT}$, then one can
rephrase the unification of the gauge couplings (a bottom-up result)
as a prediction for the weak mixing angle ($\sin^2\theta_W$) and the
strong coupling ($\alpha_3$) at the electroweak scale (a top-down result).
The larger gauge symmetry is accompanied by new gauge bosons (which conspire to
yield a single unified gauge coupling above $M_{\rm GUT}$) and new Higgs bosons
to effect the gauge symmetry breaking down to the Standard Model gauge group.
Typical unified groups include $SU(5)$, $SO(10)$, and ``flipped" $SU(5)\times
U(1)$. The new GUT degrees of freedom have
various observable consequences. One of these is the violation of baryon
number, which leads to the decay of protons with a lifetime $\tau_p\gsim
10^{32}\y$ through various channels such as $p\to e^+\pi^0$ and
$p\to K^+\bar\nu$. The unified symmetry implies that the Standard Model
particles
are grouped into larger representations, and thus their properties are
correlated. These correlations lead, for example, to an explanation of the
charge quantization observed in the Standard Model, and to relations
between the Yukawa couplings (\eg, $\lambda_b=\lambda_\tau$).
Another interesting consequence of the existence of a large mass scale
and of our calculability assumption, is the need to restrict the Standard
Model top-quark Yukawa coupling ($\lambda_t\lsim1$) so that it does not blow up
below the
unification scale. This constraint implies $m_t\lsim200\GeV$.

\subsection{Supergravity and superstrings} 
Despite their successes, supersymmetric GUTs provide only relations
among the Standard Model parameters, with no possibility of
first-principles calculations of their actual values. (Of course, GUTs predict
new phenomena beyond the Standard Model so that they can themselves
be in principle tested.) Among the uncalculated parameters we have
the many supersymmetric particle masses. These, in fact, are crucial to
the unification program, as only if they are in the
${\cal O}(100\GeV)-{\cal O}(1\TeV)$ range does one achieve unification. The
needed breaking of supersymmetry is phenomenologically viable only if
supersymmetry is a {\em local} symmetry (as the gauge symmetries are). Local
supersymmetry necessarily involves gravitational interactions, and thus the
name {\em supergravity}. In a supergravity theory the 
supersymmetry-breaking parameters are calculable in terms of just two
functions: the K\"ahler function and the gauge kinetic function. Thus
a great synthesis of the unknown parameters can be attained if these
functions are known or can be somehow parametrized. In fact, one
of the simplest possible choices for the K\"ahler function gives
{\em universal} scalar masses, \ie, the masses of the scalar 
superpartners (squarks, sleptons, etc) are all equal at the Planck scale.
(Renormalization group evolution down to low energies breaks this
degeneracy.) These ``minimal" models can be described in
terms of just {\em four} parameters, and have received a great
deal of attention in the literature ever since their inception, and
especially since the advent of LEP.

A special and interesting class of supergravity models aims at
solving two thorny problems in supersymmetry: the vanishing of
the cosmological constant and the ultimate determination of
the scale of the supersymmetric spectrum. In these {\em no-scale}
supergravity models, the tree-level cosmological constant
vanishes, and the scalar potential possesses a flat direction
along which the scale of the supersymmetric spectrum (the
gravitino mass $m_{3/2}$) is undetermined.  At the electroweak
scale this flat direction is ``bent" and the gravitino naturally
acquires a mass of electroweak size. This mechanism satisfies
automatically our naturalness assumption.

Another consequence of supergravity is the {\em radiative electroweak
symmetry breaking mechanism}. The Higgs mass-squared parameter in the
scalar potential, starting from a positive value at the Planck scale, decreases
in magnitude as the scale is lowered, eventually vanishing and turning
negative, signalling the dynamical breaking of the electroweak symmetry. In
contrast, in the Standard Model the negative Higgs mass-squared parameter is
put in by hand. The radiative breaking mechanism relies on the running of the
mass parameters and assures that the mass-squared of charged and colored
particles remain positive. Even more interesting is the fact that this
phenomenon is possible only in the presence of a not-too-light top quark 
($m_t\gsim60\GeV$). 

Supergravity theories provide an effective description of
quantum gravity, valid at scales $Q<M_{Pl}$. As such they are
non-renormalizable and contain an infinite number of 
higher-dimensional operators suppressed by powers of the
Planck mass. To make further progress in our quest for
understanding the parameters of the Standard Model
(or the MSSM), we need to  compute the K\"ahler
function and gauge kinetic function, which determine the
spectrum of supersymmetric particles. We also need  to
compute the Yukawa couplings which give rise to the Standard
Model fermion masses and quark mixing angles. The ``prototheory"
that we seek must in fact have no free parameters. It turns out
that the solution to our problems is offered by {\em string theory},
wherein elementary particles are replaced by tiny strings of
dimension $\ell_{Pl}\sim10^{-33}\cm$. In string theory {\em all}
physical parameters are in principle calculable in terms of
the Planck mass, or ratios of the Planck mass to other dynamically
determined mass scales (such as the vacuum expectation value
of the dilaton field, which determines the string gauge coupling).
As is well-known, at the present stage of its development string theory
has a very large number of equivalent ground states, or
``string models". Each of these models has a continuously
connected family associated with it, parametrized by fields with
flat potentials called {\em moduli}. Because of the present
inability to pinpoint {\em the} string vacuum (if it is indeed unique),
it is widely perceived that string theory can make no predictions.
This perception is incorrect. For one thing, in string theory one knows
that the gauge couplings become unified at the string scale $M_{\rm string}
\sim10^{18}\GeV$, irrespective of the existence of a unified gauge
group at this scale. Also, if one focuses on any given string model,
in principle all model parameters are explicitly calculable,
in particular the supergravity K\"ahler and gauge kinetic functions
and the Yukawa couplings. Thus, in specific string models one should
be able to calculate {\em all} of the Standard Model and
MSSM parameters. Consequences of the unparalleled calculability
of string models include that the supersymmetry-breaking
parameters should typically be {\em not} universal, and 
that the top-quark mass should be large $m_t\sim(150-200)\GeV$.

\subsection{Observable supersymmetry}
Another feature of many supersymmetric models
is the existence of a discrete {\em R-parity} symmetry which implies 
that real supersymmetric particles are always produced in pairs. That is,
they cannot appear as intermediate (virtual) states in tree-level processes
involving
only external Standard Model particles (\eg, in $e^+e^-\to\mu^+\mu^-$).
They can first appear in such processes at the loop level, and therefore their
indirect
effects are naturally suppressed. R-parity also implies the existence of
a new stable particle: the {\em lightest supersymmetric particle} (LSP). This
particle has important cosmological consequences, since in large
regions of parameter space it can account for the all-pervading (cold)
dark matter in the Universe.

Supersymmetric particles are being actively searched
for at several experimental facilities. At the Tevatron 
the strongly-interacting gluinos and squarks have been the long-time
target in the $p\bar p$ collisions, but the kinematical reach of the collider
has been nearly reached. This has prompted the search for particles with
feebler interaction strengths, but which could still be produced because of 
their lighter masses, such as the weakly-interacting gauginos. A proposed
high-luminosity upgrade of the Tevatron would be particularly sensitive to
the gaugino signal. At LEP~1 (1989--1995), the
large number of $Z$ bosons produced ($\sim20$ million) has ruled out
the existence of new particles with unsuppressed couplings to the $Z$
and with masses below $\sim{1\over2}M_Z$.  This result has constrained the
masses
of most supersymmetric particles. The search for the Higgs boson at LEP~1
($m_H\gsim65\GeV$) has also constrained the Higgs bosons of supersymmetric
models. The LEP energy upgrade (LEP~2) should push these limits even
further starting in 1996. More realistically, CERN's Large Hadron Collider
(LHC) (ca.~2004)
should provide definitive evidence for low-energy supersymmetry or,
if no supersymmetric particles are observed, render it unappealing. Once
supersymmetry is established, a dedicated electron-positron linear collider
would be ideal for what has been called ``sparticle spectroscopy."

Indirect searches for supersymmetric particles are a useful 
complement to direct searches. The recently observed $b\to s\gamma$
loop process at Cornell's electron-positron storage ring by the CLEO
Collaboration, puts important constraints on the supersymmetric
parameter space. The upcoming anomalous magnetic moment of the muon
($g$--2) experiment at Brookhaven
should also prove to be a  stringent test of supersymmetric models. 
Finally, as the presumed main component of the galactic dark matter halo, the
LSP is being searched for in various direct and indirect ways.

\subsection{Disclaimer}
Because of the intended nature of this review many details have been left out,
and particular emphasis has been placed on developments that have occurred
over the last decade. The reader is encouraged to consult the standard accounts
\cite{OldReviews}, as well as a number of recent detailed reviews on the
subjects of supersymmetric model-building \cite{DM}, supersymmetry
phenomenology \cite{Baer}, and supersymmetric dark matter \cite{Griest}. I
should also point out that due to lack of space and expertise, the more formal
topics of supersymmetric theories have been left uncovered. A particularly
glaring omission pertains to the large and relatively recent body of literature
on exact results in supersymmetric gauge theories and the rapidly evolving
topic of superstring dualities; for recent reviews see Refs.~\cite{Seiberg} and
\cite{Schwarz} respectively.

\section{Low-energy Supersymmetry}
\label{sec:MSSM}
\subsection{Supermultiplets}
As discussed in Sec.~\ref{sec:Intro} (see Fig.~\ref{fig:BigPicture}), our first
step away from the Standard Model consists of introducing low-energy
supersymmetry. Supersymmetry is a space-time symmetry, as are the well-known
Lorentz and Poincar\'e symmetries. Under the Poincar\'e group particles
are labelled by their mass and their spin. The mass can be any positive real
number or zero, whereas the spin must be integer or half-integer. These
space-time symmetries are distinct from internal symmetries (such as gauge
symmetries), which do not change the mass or spin of a particle but can change
its internal quantum numbers (such as electric or color charge). Conversely,
the space-time symmetries do not change the internal quantum numbers of a
particle. One says that space-time symmetries ``commute" with internal
symmetries. Supersymmetry, as a space-time symmetry, changes the spin of a
particle (but not its mass) and leaves all its internal quantum numbers
unchanged. There are various kinds of supersymmetric theories, depending on the
number of different supersymmetry {\em generators}. Each of these generators
can change the spin of a particle by half-a-unit. In building realistic models
one only considers $N=1$ supersymmetry, \ie, models with a single supersymmetry
generator. The phenomenological problem of extended ($N\ge2$) supersymmetries
is that fermions with different chiralities get related by supersymmetry, a
result which is incompatible with the left-handed nature of the electroweak
interactions. Two kinds of {\em supermultiplets} are mostly used in building
supersymmetric models: the {\em chiral supermultiplet} contains a chiral
spin-$1\over2$ fermion and a spin-0 scalar, whereas the {\em vector
supermultiplet} contains a spin-1 vector boson and a Majorana spin-$1\over2$
fermion. These supermultiplets can accommodate all of the Standard Model
particles and their superpartners.

\subsection{The Minimal Supersymmetric Standard Model}
The Minimal Supersymmetric Standard Model (MSSM) \cite{MSSM} is generally 
considered to be the closest one can get to the Standard
Model, if one allows for the possibility of low-energy supersymmetry. This
model is very general but has little predictive power, with more than
30 parameters required to fully describe it. Nonetheless, it should
contain any model obtained from the more constrained theories
that we describe in the following sections. In the MSSM, each Standard
Model particle is paired with a superpartner. The fermions (quarks
and leptons) belong in chiral supermultiplets together with the spin-0
sfermions (squarks and sleptons), whereas the gauge bosons
(photon, gluon, $W,Z$)
belong in vector multiplets together with the Majorana spin-$1\over2$
gauginos (photino, gluino, wino, zino). Finally, the Higgs boson
is paired with a spin-$1\over2$ Higgsino in a chiral multiplet. The
novelty is that two Higgs doublets are {\em required} in
supersymmetric models, as we discuss shortly. The MSSM
particle content and notation are collected in Table~\ref{Table1}.

\begin{table}[t]
\caption{Field content and notation in the Minimal Supersymmetric Standard
Model (MSSM). Arrows indicate fields that mix due to the Yukawa interactions,
and the corresponding physical fields that result.}
\label{Table1}
\bigskip
\hrule
\begin{center}
\begin{tabular}{lccc||lccccc}
\underline{Quarks}&&&&\quad \underline{Squarks}&&&&\\
(spin-$1\over2$)&$\left({u\atop d}\right)_L$&$u_R$&$d_R$&
\quad(spin-0)&$\left({\tilde u\atop\tilde d}\right)_L$
&$\tilde u_R$&$\tilde d_R$&&\\
&$\left({c\atop s}\right)_L$&$c_R$&$s_R$&
&$\left({\tilde c\atop\tilde s}\right)_L$&$\tilde c_R$&$\tilde s_R$&&\\
&$\left({t\atop b}\right)_L$&$t_R$&$b_R$&
&$\left({\tilde t\atop\tilde b}\right)_L$&$\tilde t_R$&$\tilde b_R$
&$\longrightarrow$&$\tilde t_{1,2}\,,\tilde b_{1,2}$\\
&&&&&&&&\\
\underline{Leptons}&&&&\quad \underline{Sleptons}&&&&\\
(spin-$1\over2$)&$\left({e\atop \nu_e}\right)_L$&$e_R$&&
\quad(spin-0)&$\left({\tilde e\atop\tilde \nu_e}\right)_L$&$\tilde e_R$&&&\\
&$\left({\mu\atop \nu_\mu}\right)_L$&$\mu_R$&&
&$\left({\tilde \mu\atop\tilde \nu_\mu}\right)_L$&$\tilde \mu_R$&&&\\
&$\left({\tau\atop\nu_\tau}\right)_L$&$\tau_R$&&
&$\left({\tilde \tau\atop\tilde \nu_\tau}\right)_L$&$\tilde t_R$&
&$\longrightarrow$&$\tilde \tau_{1,2}$\\
&&&&&&&&\\
\underline{Gauge bosons}&&&&\quad\underline{Gauginos}&&&&\\
(spin-1)&$g$&&&\quad(spin-$1\over2$)&$\tilde g$\\
&$\gamma$&&&&$\tilde\gamma$&&&\multicolumn{2}{l}{Neutralinos}\\
&$Z$&&&&$\widetilde Z$&&&$\longrightarrow$&$\chi^0_{1,2,3,4}$\\
&$W^\pm$&&&&$\widetilde W^\pm$&&&
\multicolumn{2}{l}
{$\{\tilde\gamma,\widetilde Z,\widetilde H^0_{1,2}\}$}\\
&&&&&&&&\\
\underline{Higgs bosons}&&&&\quad
\underline{Higgsinos}&&&&\multicolumn{2}{l}{Charginos}\\
(spin-0)&$h,H,A$&&&\quad(spin-$1\over2$)&$\widetilde H^0_{1,2}$
&&&$\longrightarrow$&$\chi^\pm_{1,2}$\\
&$H^\pm$&&&&$\widetilde H^\pm$&&&
\multicolumn{2}{l}
{$\{\widetilde W^\pm,\widetilde H^\pm\}$}
\end{tabular}
\bigskip
\hrule
\end{center}
\end{table}

Since supersymmetry commutes with the gauge symmetry of the
Standard Model, the MSSM is still an $SU(3)_C\times SU(2)_L\times
U(1)_Y$ gauge theory. That is, the gauge interactions of the
superpartners are the {\em same} as those of the ordinary
particles. For instance, the left-handed $SU(2)$ fermion doublet
$\left({e\atop\nu_e}\right)_L$ is partnered
with the scalar doublet 
$\left({\tilde e\atop\tilde\nu_e}\right)_L$.
The Feynman rules for the superpartners allow the same
interactions, with the same strength, although one must take into
account the different spinor nature of the particles.

An important assumption usually made in the MSSM is the imposition of
a discrete ``R-parity" symmetry, that assigns a charge of $-1$ to
the superparticles, and a charge of 0 to the regular particles (and the
Higgs bosons). This symmetry restricts the possible interactions in
the theory: at every interaction vertex there must be an {\em even} number
(0,2,4) of superparticles. One important consequence of this asssumption is
the elimination of possible interactions, allowed by the gauge symmetry, that
would lead to very fast proton decay (via dimension-four operators). Another
important phenomenological consequence
is that superparticles cannot appear as intermediate states in tree-level
processes involving regular external particles only, thus ``protecting"
the Standard Model predictions to lowest order. Since superparticles are
known to be not-too-light, their first tree-level appearance requires colliders
with sufficient energy to pair produce them. A related consequence is that
there should exist a lightest supersymmetric particle (LSP), which is
absolutely stable, and which should appear copiously in any reaction that
produces superparticles. Astrophysical considerations restrict the possible LSP
choices to neutral and colorless particles \cite{EHNOS}, \ie, the lightest
neutralino ($\chi$) or the sneutrino. For various theoretical and
phenomenological reasons \cite{DM}, it is the neutralino which is usually
associated with the LSP. Supersymmetric models without R-parity have also been
considered in the literature \cite{Rparityx}.

\subsection{Yukawa and scalar interactions}
The Yukawa interactions and the scalar potential in supersymmetry
are more constrained than in the Standard Model. The
Yukawa interactions derive from an object called the
superpotential ($W$), which is a cubic function of the
chiral superfields, with only one chirality allowed for all
superfields present. One says that $W$ is a {\em holomorphic}
function of its arguments.\footnote{This property has played a key role
in recent developments concerning exact  results in supersymmetric gauge
theories \cite{Seiberg}.} The superpotential in the MSSM (restricted by
R-parity) contains the Yukawa couplings giving rise to the fermion masses, \ie,
\begin{equation}
W=\lambda^{ij}_u Q_i u^c_j H_2+\lambda^{ij}_d Q_i d^c_j H_1
+\lambda^{ij}_e L_i e^c_j H_1+\mu H_1 H_2\ .
\label{eq:W}
\end{equation}
Note that in the MSSM two Higgs doublets are {\em required} to provide all the
needed Yukawa couplings since, in contrast with the Standard Model, the
superpotential does not allow conjugate fields. (Next-to-minimal supersymmetric
models typically include an additional Higgs singlet field in the low-energy
spectrum \cite{NMSSM}.) The sum of the squares of the two Higgs vacuum
expectation values (vevs) is constrained by the usual Higgs vev in the Standard
Model. However, their ratio is undetermined, and it is denoted by the parameter
$\tan\beta=\VEV{H_2}/\VEV{H_1}$.

The scalar potential in the MSSM contains both
supersymmetry-conserving and supersymmetry-breaking terms. The
supersymmetry-conserving terms come from two sources, schematically
\begin{equation}
V_{\rm susy}\sim \sum_i(g\phi^*_i\phi_i)^2+\sum_i\left|{\partial
W\over\partial\phi_i}\right|^2\ ,
\label{eq:Vsusy}
\end{equation}
where the sum runs over all scalar fields $\phi_i$ in the model.
The first contribution (the ``D-terms") provide the gauge and Higgs boson
masses and interactions. These terms are analogous to their Standard Model
counterparts, with one important exception. In the MSSM the Standard Model
quartic Higgs coupling ($\lambda$) is predicted to be $\lambda\sim g^2$.
The second contribution (the ``F-terms") provide the quartic scalar
interactions among the squarks, sleptons, and Higgs bosons, and provide an
additional contribution to the Higgs boson mass matrix from the ``$\mu$ term"
in Eq.~(\ref{eq:W}). The supersymmetry-breaking contributions to the scalar
potential
\begin{equation}
V_{\rm soft}\sim \sum_i m^2_i|\phi|^2 + A W_3+B W_2\ ,
\label{eq:Vsoft}
\end{equation}
provide masses to the squarks, sleptons, and Higgs bosons, as well as a set of
trilinear ($AW_3$) and bilinear ($BW_2$) interactions mimicking the trilinear
($W_3$) and bilinear ($W_2$) terms in the superpotential. The theory also
includes supersymmetry-breaking masses for the gauginos: $m_{\tilde g}$ for the
gluino, $M_2$ for the wino, and $M_1$ for the bino.

Because of the required two Higgs-boson doublets, the Higgs boson spectrum
in the MSSM is much richer than in the Standard Model: two neutral scalars
($h,H$), a neutral pseudoscalar ($A$), and a charged scalar ($H^\pm$). One
can show that in the MSSM the $h$ Higgs boson is always light:  $m_h^{\rm
tree}<M_Z|\cos2\beta|$. The upper limit is approached when the supersymmetry
breaking contributions in $V_{\rm soft}$ are large, in which case the
remaining Higgs bosons become very heavy and decouple. Moreover, the
interactions of the light Higgs boson become indistinguishable from the Higgs
boson in the Standard Model. One of the most phenomenologically
relevant realizations of the last several years is that one-loop corrections
to the Higgs boson mass in supersymmetric models are enhanced by a heavy
top quark \cite{ERZ},
\begin{equation}
(m^2_h)^{\rm one-loop}\sim (m^2_h)^{\rm tree}
+ c (m^4_t/m^2_Z)\ln(m^2_{\tilde t}/m^2_t)\ .
\label{eq:mhloop}
\end{equation}
For $m_t\sim180\GeV$, the upper limit becomes $m_h\lsim130\GeV$, greatly
affecting the discovery potential for the Higgs boson at LEP.

\subsection{Experimental limits}
Experimental limits on the MSSM model parameters had been rather mild before
the advent of the Tevatron and LEP. At the Tevatron the strongly-interacting
squarks and gluino can be pair produced ($p\bar p\to \tilde g\tilde g,
\tilde q\tilde q,\tilde g\tilde q$) with sizeable cross sections. After
production these particles decay in a ``cascade" until the LSP is reached.
Since the LSP is ``invisible", it leads to an imbalance in the
transverse-momentum total sum, or in practice to a ``missing transverse energy"
(missing $E_T$) signature. The latest experimental limits
from the Tevatron indicate that \cite{TeVsqg}
\begin{equation}
m_{\tilde q}\gsim175\GeV\,,\quad m_{\tilde g} \gsim175\GeV\,;\qquad
m_{\tilde q}\approx m_{\tilde g} \gsim230\GeV\ .
\label{eq:sqglimits}
\end{equation}
LEP~1 limits on superparticle masses are generally close to ${1\over2}M_Z$
for all pair-produced particles with unsuppressed couplings to the $Z$
boson \cite{LEP1Limits}, this includes the squarks, sleptons, and charginos.
The neutralinos, which are admixtures of the photino, zino, and neutral
Higgsinos, can couple to the $Z$ boson only through their Higgsino components.
In general, the masses and compositions of the neutralinos depend on the
choices for $\mu$, $M_1,M_2$, and $\tan\beta$. The LEP~1 limits on neutralinos
therefore depend on these three parameters \cite{LEP1chi0}. In experimental
analyses of the MSSM, a ``GUT assumption" is often made, which relates the
masses of the gauginos to one another
\begin{equation}
M_1=\coeff{5}{3}\tan^2\theta_W\, M_2\,,\qquad
M_2=\left({\alpha_2\over\alpha_3}\right)m_{\tilde g}\ .
\label{eq:GUTassumption}
\end{equation}
As we discuss below, this result in fact follows in many GUTs, but in the MSSM
it just serves as a simplifying assumption. Assuming this relation one can use
the experimental limits on $m_{\tilde g}$ to bound $M_1,M_2$ and therefore the
lightest neutralino mass: $m_{\chi^0_1}\gsim20\GeV$ \cite{LEP1chi0}. If the GUT
assumption is {\em not} made, very light neutralinos are still allowed,
provided that they be mostly higgsino admixtures (this requires $\mu$
to be small too) \cite{Feng}. The Higgs bosons have also been searched for
at LEP~1, most notably
the lightest one via $e^+e^-\to Z\to Z^* h\to f\bar f h$. The production cross
section differs from the Standard Model one only by an overall factor of
$\sin^2(\alpha-\beta)$, where $\alpha$ is the Higgs mixing angle. For arbitrary
values of this factor, and assuming the dominant Standard Model $h\to b\bar b$
decay, LEP~1 has determined that $m_h\gsim40\GeV$ \cite{LEP1Higgs}. The
corresponding limit in the Standard Model is $m_{H_{\rm SM}}>65\GeV$
\cite{LEP1Higgs}. This limit is actually applicable in the MSSM if the
superparticle masses are large, since in this limit the lightest supersymmetric
Higgs boson becomes indistinguishable from the Standard Model Higgs boson.

\subsection{Shortcomings}
The MSSM has many parameters, and therefore a wide range of possible
experimental predictions, \ie, very little predictive power. In
addition to this obvious shortcoming, there are more subtle hints that make the
need to go beyond the MSSM pressing. In the Standard Model, flavor-changing
neutral currents (FCNC) are absent at tree-level and are sufficiently
suppressed in one-loop processes (such as $K-\bar K$ mixing and $\mu\to
e\gamma$) because of the small light-fermion masses (\eg,
$(m^2_c-m^2_u)/M^2_W\ll1$). Such processes receive new one-loop contributions
in the MSSM, with the analogous requirement being $(m^2_{\tilde c}-m^2_{\tilde
u})/m^2_{\tilde q}\ll1$, and similarly for the sleptons. In the same vein, new
supersymmetric phases contribute to the dipole moment of the neutron, and are
experimentally required to be rather suppressed ($\phi_{\rm susy}\lsim10^{-3}$)
unless the supersymmetric spectrum is rather heavy \cite{dn}. Such stringent
constraints must be imposed by hand on the MSSM parameter space.

\section{Supersymmetric Grand Unification}
\label{sec:GUTs}
\subsection{Running gauge couplings}
The most basic prediction of a grand unified theory (GUT) that is to encompass
the Standard Model, is that the low-energy gauge couplings should converge
to a single value at a sufficiently high mass scale ($M_{\rm GUT}$).
Above this scale new degrees of freedom are excited, and the gauge couplings
do not diverge again, but continue as a single unified coupling. From the
low-energy point of view (the bottom-up approach), the convergence test
must be passed. From the high-energy point of view (the top-down approach),
there is a distinct prediction for, \eg, $\sin^2\theta_W$ which, when evolved
down to low-energies, should agree with the experimental measurement. As early
as 1987 \cite{amaldi-costa}, the experimental determinations of the low-energy
gauge couplings were precise enough to indicate that, even though both the
Standard Model and the MSSM passed the convergence test, the MSSM did so more
persuasively -- in retrospect, the Standard Model had actually failed, but it
managed to hide behind the experimental uncertainties. When the LEP data became
available starting in 1989, the failure of the Standard Model as a unified
theory became clear, as did the success of the MSSM \cite{EKN}. The reason for
the different outcomes in the Standard Model versus the MSSM is not the gauge
symmetry, but rather the supersymmetry and the spectrum of light particles.
Because of this ambiguity, schemes have been devised to ``fix" the running of
the gauge couplings in the Standard Model by adding ad-hoc intermediate-scale
particles \cite{Frampton}. These schemes may be logically viable but, without
supersymmetry, are incapable of solving the gauge hierarchy problem.

The running of the gauge couplings in unified models is familiar from our
previous experience with the running coupling ``constant" in QCD. The basic
quantity of interest is the {\em beta function}, which quantifies the rate of
change of the gauge coupling with the logarithm of the mass scale ($Q$), and
depends on the numbers of light particles in the given model (at a given
scale). Writing this differential equation in terms of the inverses of the
``structure constants" $\alpha_i=g^2_i/4\pi$, to next-to-leading order one
obtains
\begin{equation}
{d\alpha^{-1}_i\over dt}=-{b_i\over2\pi}
-{1\over8\pi^2}\sum^3_{j=1}b_{ij}\alpha_j\ ,
\label{eq:gRGE}
\end{equation}
where $i=Y,2,3$ and $t=\ln(Q/M_{\rm GUT})$.  In this form it is evident that
(to lowest order) the convergence of the gauge couplings reduces to the meeting
of three lines [$\alpha^{-1}_i(t)$] on a plane. In the Standard Model and in
the MSSM, the one-loop ($b_i$) beta functions are given by \cite{EKN}
\begin{equation}
b^{\rm SM}_i=\left(\coeff{41}{10},-\coeff{19}{6},-7\right)\,;\qquad
b^{\rm MSSM}_i=\left(\coeff{33}{5},1,-3\right)\,.
\label{eq:betas}
\end{equation}
The relative increase of the beta functions in the MSSM versus the SM is
expected because of the larger number of fermions in the MSSM (\ie, the
gauginos). The MSSM (SM) beta functions are valid for mass scales above (below)
the masses of {\em all} the particles in the MSSM. In solving the RGEs in
Eq.~(\ref{eq:gRGE}) between the scales $Q=M_{\rm GUT}$ and $Q=M_Z$, in first
approximation one assumes that all MSSM particles have the same mass (\eg,
$M_Z$), and that one-loop beta functions suffice. In the next level of
approximation one includes two-loop corrections to the beta functions
($b_{ij}$), and accounts for the many ``thresholds" at which the supersymmetric
particles decouple. This is a complicated exercise which has been performed in
several different schemes \cite{OldGUTs,NewGUTs}. The original motivation for
these calculational refinements was to gain some information about the spectrum
of supersymmetric particle masses by fitting it to get the ``best" convergence
at $M_{\rm GUT}$. This exercise proved to be futile because of the very nature
of the unification process, that itself involves thresholds of much less
understood particles near the GUT scale \cite{OldGUTs}.

\begin{figure}[t]
\vspace{5in}
\includegraphics{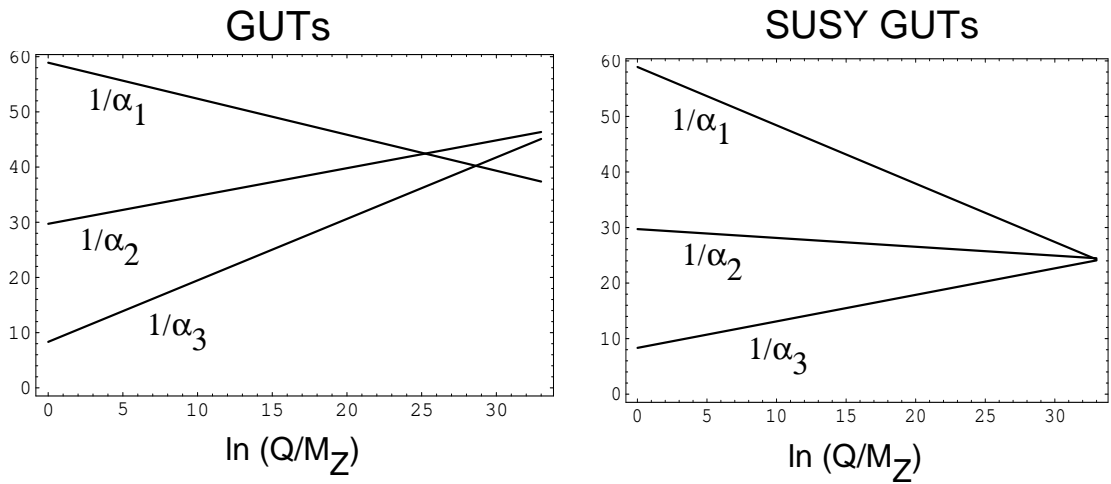}
\vspace{-2in}
\caption{A lowest-order comparison of the convergence of the gauge couplings in
the Standard Model versus the MSSM. The convergence of the gauge couplings is a
necessary but not a sufficient requirement for the existence of a GUT.}
\label{fig:Runnings}
\end{figure}

The gauge coupling RGEs in Eq.~(\ref{eq:gRGE}) are supplemented with the
initial conditions
\begin{eqnarray}
\alpha^{-1}_e(M_Z)&=&127.9\pm0.1,\\
\alpha_3(M_Z)&=&0.118\pm0.006,\\
\sin^2\theta_W(M_Z)&=&0.23186\pm0.00034.\label{eq:sin2exp}
\end{eqnarray}
The $U(1)_Y$ and $SU(2)_L$ gauge couplings are related to these by\footnote{The
factor of $5\over3$ in the definition of $\alpha_1$ follows from a rescaling
of the Standard Model hypercharges so that, once embedded in a simple gauge
group, all generators are equally normalized. From $Q=T_3+Y=T_3+cY'$,
$g_1=cg'_1$, one determines $c$ such that $\Tr T^2_3=\Tr Y'^2$ over {\em any}
representation of the GUT group. In $SU(5)$ $c=\sqrt{5/3}$, and thus
$\alpha_1\propto g^2_1\propto {5\over3}{g'}_1^2$.}
$\alpha_1={5\over3}(\alpha_e/\cos^2\theta_W)$ and
$\alpha_2=(\alpha_e/\sin^2\theta_W)$. In Fig.~\ref{fig:Runnings} we show a
lowest-order comparison of the convergence of the gauge couplings in the MSSM
versus the SM.

\subsection{SU(5) GUTs}
The convergence of the gauge couplings, as observed from the bottom-up
approach, suggests that at scales $Q=M_{\rm GUT}\sim10^{16}\GeV$, a new
structure emerges, that of a unified theory. Let us take $SU(5)$ as a prototype
grand unified theory \cite{GG}, and discuss various features that are
qualitatively common to all unified theories. Later we compare
$SU(5)$ to other unified groups, such as $SO(10)$ and $SU(5)\times U(1)$.
In $SU(5)$, the Standard Model fermions of {\em each} generation are
accommodated in the \rb{5} and \r{10} representations:
\begin{equation}
\bar{\bf5}=(\bar{\bf3},1)+(1,{\bf2})=\{d^c,L\}\,,\qquad
{\bf10}=({\bf3},{\bf2})+(\bar{\bf3},1)+(1,1)=\{Q,u^c,e^c\}\ .
\label{eq:5bar+10}
\end{equation}
This combination is anomaly-free. Because the right-handed down quark ($d^c$)
and the lepton doublet ($L$) belong to the same $SU(5)$ representation, and the
$SU(5)$ group generators are traceless (in particular the electromagnetic
charge), the {\em charge quantization} relation $3q_{d^c}+q_e=0$ follows for
each generation. The gauge bosons of $SU(5)$ belong to the adjoint (\r{24})
representation,
\begin{equation}
{\bf24}=({\bf8},1)+(1,{\bf3})+(1,1)+({\bf3},{\bf2})+(\bar{\bf3},{\bf2})
=\{g,W^\pm,W^0,B,X^{\pm4/3},Y^{\pm1/3}\}
\label{eq:24}
\end{equation}
and contain the 12 Standard Model gauge bosons plus 12 new, heavy ($\sim M_{\rm
GUT}$), charged, colored gauge bosons denoted by $X,Y$. The breaking of the
unified symmetry down to the Standard Model gauge group is effected by Higgs
bosons in the \r{24} representation, whose single neutral component [the
$(1,1)$ in Eq.~(\ref{eq:24})] gets a vev. The real parts of
the Higgs fields in the $({\bf3},{\bf2})+(\bar{\bf3},{\bf2})$ representations
are eaten by the $X,Y$ gauge bosons, whereas the imaginary parts acquire their
same mass ($M_V$). Further Higgs bosons belong to the $({\bf8},1)$,
$(1,{\bf3})$, and $(1,1)$ representations with masses
$M_\Sigma,M_\Sigma,{1\over5}M_\Sigma$. The Higgs-boson doublet in the Standard
Model is promoted to a Higgs pentaplet $H=\{H_2,H_3\}$, which contains a new,
colored Higgs triplet field with mass $M_{H_3}$. The mass of the Higgs triplet
is constrained by proton decay limits (see below) to be no smaller than the GUT
scale, \ie, $M_{H_2}/M_{H_3}\sim M_Z/M_{\rm GUT}\ll1$. This {\em
doublet-triplet splitting} problem is resolved in minimal $SU(5)$ by resorting
to a severe fine tuning \cite{MSSM}. An alternative $SU(5)$ GUT model, the {\em
missing doublet model} (MDM) naturally solves this problem \cite{MPM,MDM},
although at the expense of introducing new \r{50},\rb{50} Higgs multiplets and
a \r{75} to break the gauge symmetry. Further alternative mechanisms to address
this problem have also been suggested \cite{PGBM}. Since the three mass
parameters ($M_V,M_\Sigma,M_{H_3}$) in minimal $SU(5)$ (or even more parameters
in the MDM) are generally not all equal, one encounters a ``heavy threshold"
behavior near $M_{\rm GUT}$, which smears the (lowest-order) concept of a
single unification point.

{}From the GUT (or top-down) approach, one obtains an expression for the weak
mixing angle: $\sin^2\theta_W(M_{\rm GUT})=3/8$. The test of the unified
theory is the prediction for $\sin^2\theta_W$ at the electroweak scale;
to lowest order:
\begin{eqnarray}
\sin^2\theta_W(M_Z)|_{\rm GUTs}&=&
{1\over6}+{5\over9}{\alpha\over\alpha_3}\approx0.203\ ;
\label{eq:sin2GUTs}\\
\sin^2\theta_W(M_Z)|_{\rm SUSY GUTs}&=&
{1\over5}+{7\over15}{\alpha\over\alpha_3}\approx0.230\ .
\label{eq:sin2SUSYGUTs}
\end{eqnarray}
The difference between these two predictions, even though small, is far
greater than the present uncertainty in the experimental determination
of $\sin^2\theta_W(M_Z)$ (\ref{eq:sin2exp}). In the state-of-the-art
calculations one inputs the precisely measured value of $\sin^2\theta_W$,
and obtains a prediction for the not-as-well-measured strong coupling
[$\alpha_3(M_Z)$]. This is done by taking into account a variety of subleading
effects, such as two-loop contributions to the beta functions, effect of Yukawa
couplings, effect of finite (non--leading-logarithmic) corrections, light
supersymmetric particle thresholds, heavy particle thresholds, different models
of $SU(5)$ GUT physics, sophisticated decoupling methods beyond the
step-function approximation, non-renormalizable Planck-scale operators,
universal and non-universal supersymmetry-breaking masses, etc.
\cite{OldGUTs,NewGUTs,NROs}. Interestingly enough, it has been recently
realized that a more accurate treatment of the various threshold effects leads
to an increase in the predicted value of $\alpha_3(M_Z)$ \cite{NewGUTs}. In
fact, it has been suggested that minimal $SU(5)$ cannot reproduce any of the
known values of $\alpha_3(M_Z)$ (see \cite{Shifman} for a discussion of this
separate controversy), whereas the MDM can because of its richer heavy
threshold structure \cite{MDM}. In either case, the effect of GUT particles
near the Planck scale and non-renormalizable Planck-scale operators has been
emphasized as a means to resolve this impasse \cite{NROs}.

\subsection{Proton decay}
\begin{figure}[t]
\vspace{5in}
\includegraphics{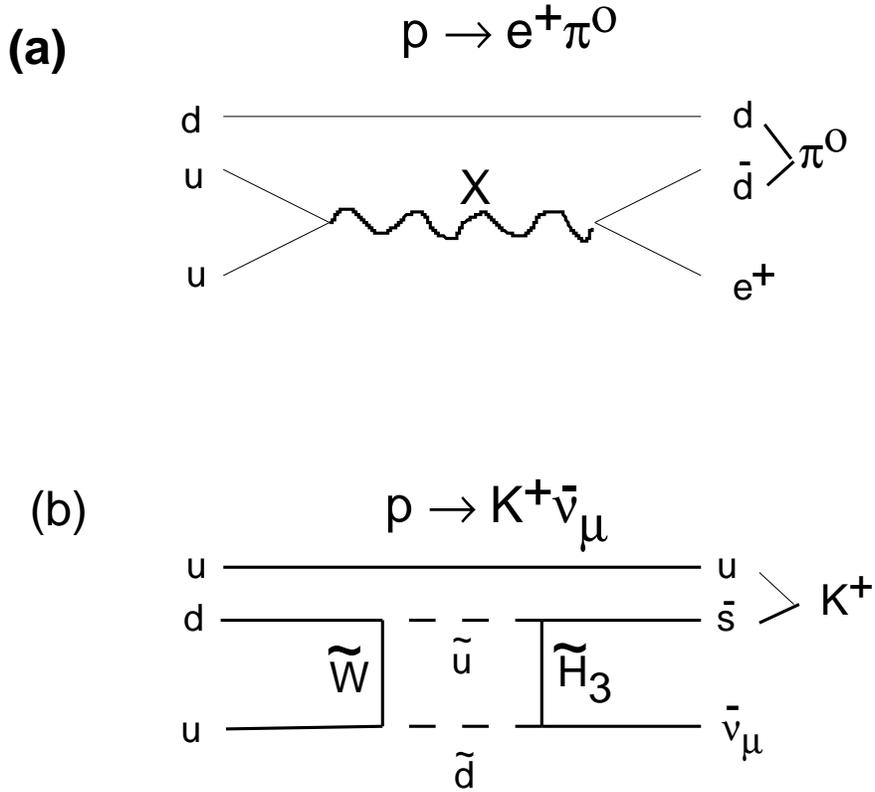}
\caption{Typical diagrams contributing to proton decay in the minimal $SU(5)$
GUT model: (a) $p\to e^+\pi^0$ mediated by the exchange of GUT gauge bosons
($X,Y$); (b) $p\to K^+\bar\nu$ mediated by the exchange of GUT Higgs bosons
($H_3$) and light supersymmetric particles.}
\label{fig:pdecay}
\end{figure}

Proton decay is perhaps the most characteristic experimental
signature of unified theories. Several modes have been searched for, most
notably $p\to e^+\pi^0$ and $p\to K^+\bar\nu$, with ``partial lifetime"
lower bounds of $5.5\times10^{32}\y$ and $1.0\times10^{32}\y$ respectively
\cite{PDG}. The forthcoming SuperKamiokande experiment should be able to probe
partial lifetimes as high as $10^{34}\y$. The $p\to e^+\pi^0$ decay channel
(see Fig.~\ref{fig:pdecay}a) is mediated by the exchange of the $X,Y$ gauge
bosons (dimension-six operators), and yields a lifetime $\tau_p\sim M^4_{\rm
GUT}/m^5_p$. From the experimental lower bound one concludes that $M_{\rm
GUT}\gsim10^{15}\GeV$, which is not in conflict SUSY GUTs predictions. However,
this lower bound disagrees with the expectation in non-supersymmetric $SU(5)$,
if we were to imagine an approximate unification scale from
Fig.~\ref{fig:Runnings}. The $p\to K^+\bar\nu_\mu$ decay channel (see
Fig.~\ref{fig:pdecay}b) arises in supersymmetric theories \cite{WSY}, with the
superpartner of the Higgs triplet (the Higgsino) mediating the conversion of
the quarks in the proton into squarks and sleptons. A loop is needed to produce
the final state particles. This is a dimension-five operator, with quadratic
dependence on the Higgs triplet mass $M_{H_3}$, and significant dependence on
the supersymmetric spectrum \cite{Oldpd}: $\tau_p\propto M^2_{H_3}
\sin^22\beta/|f|^2$,
where $f\sim m_{\chi^+_1}/m^2_{\tilde q}$ represents the one-loop dressing
function. On dimensional grounds it would appear that dimension-five operators
are much too large. However, various small factors (\ie, light-quark masses,
one-loop coefficients, etc.) make it acceptable as long as $M_{H_3}\gsim M_{\rm
GUT}$ and some important restrictions on the supersymmetric parameter space are
imposed, notably light charginos and heavy squarks \cite{Newpd}.

\subsection{Yukawa unification}
\begin{figure}[t]
\vspace{5in}
\includegraphics{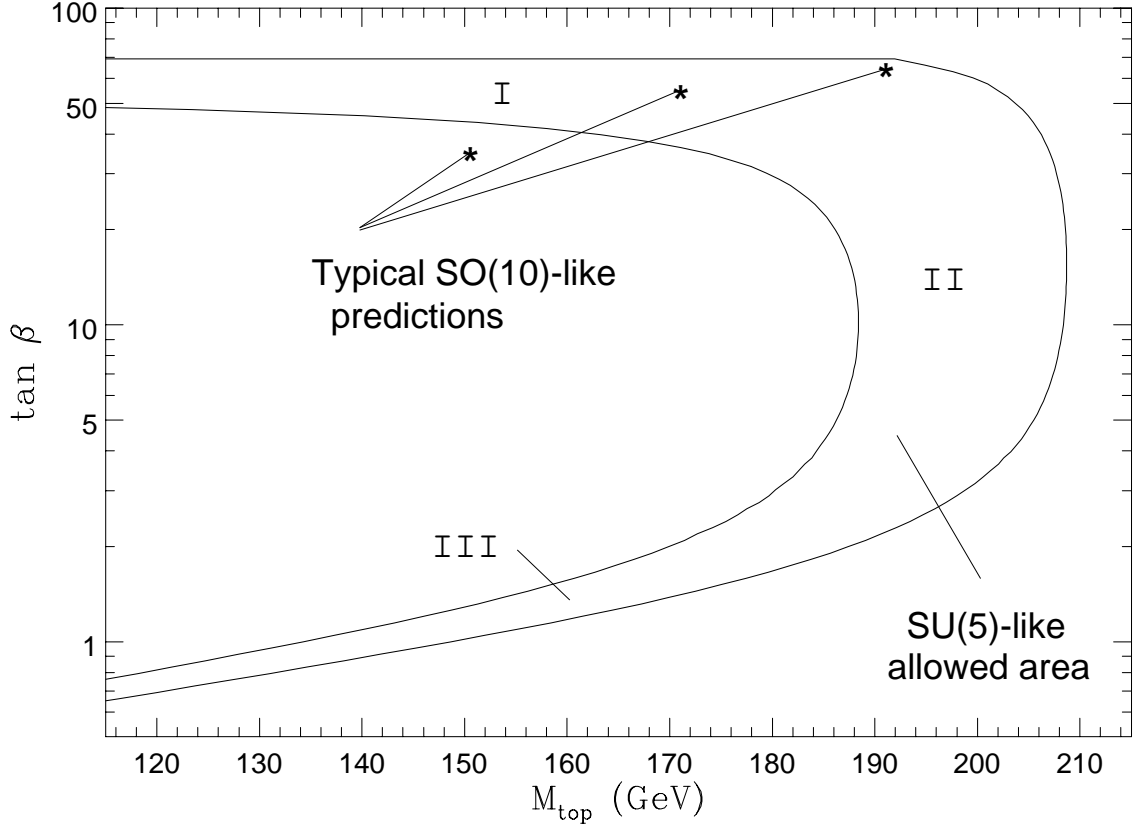}
\caption{Typical allowed area in the $(m_t,\tan\beta)$ plane from the Yukawa
unification constraints $\lambda_b=\lambda_\tau$ [`SU(5)-like'] and
$\lambda_t=\lambda_b=\lambda_\tau$ [`SO(10)-like'].}
\label{fig:KKW}
\end{figure}

Because in unified models quarks and leptons are assembled into larger
representations, one typically obtains relations among different Yukawa
couplings at the unification scale. In minimal $SU(5)$ one has two Yukawa
coupling terms
\begin{eqnarray}
\lambda_u\, {\bf10}\cdot{\bf10}\cdot H\quad&\to&\quad \lambda_u Q u^c H_2
\label{eq:up}\\
\lambda_{d,e}\, {\bf10}\cdot\bar{\bf 5}\cdot\bar H\quad&\to&\quad
\lambda_{d,e} (Q d^c H_1+L e^c H_1)
\label{eq:down,e}
\end{eqnarray}
where $H_{1,2}$ are the Higgs doublets of the MSSM. The second relation
indicates that at the GUT scale
$\lambda_b(M_{\rm GUT})=\lambda_\tau(M_{\rm GUT})$, and similarly for
the lighter generations. At low energies one has $m_b=\lambda_b(m_b) v_1$,
$m_\tau=\lambda_\tau(m_\tau)v_1$, and $m_t=\lambda_t(m_t) v_2$. Since these
parameters are interrelated, results of this analysis are usually presented as
an allowed curve (smeared by the choices of $m_b$ and $\alpha_3$) in the
$(m_t,\tan\beta)$ plane (see Fig.~\ref{fig:KKW}). As in the case of gauge
coupling unification, various sub-leading effects have been included in the
sophisticated analysis (light and heavy thresholds, possible corrections to
the $\lambda_b=\lambda_\tau$ relation, etc.) \cite{Oldbtau,btau}. The predicted
value of $m_b$ has always tended to come out uncomfortably high (\ie,
$m_b\gsim5\GeV$). Recent studies suggest that heavy supersymmetric particles in
minimal $SU(5)$, or a richer GUT structure (the MDM) help bring $m_b$ into the
preferred range \cite{BMP}.

In the $SO(10)$ GUT model \cite{OldSO10}, the Standard Model fermions are
accommodated in a single representation per generation
\begin{equation}
{\bf16}=\bar{\bf5}+{\bf10}+{\bf1}\ ,
\label{eq:16}
\end{equation}
with the novelty of a new Standard Model singlet (\r{1}) which contains a
right-handed neutrino. The Higgs pentaplets get unified into ${\bf10}=H+\bar
H$. Of the various features of $SO(10)$, let us just cite two: a {\em see-saw}
mechanism for generating small neutrino masses (typically
$m_\nu\sim m^2_u/M$, where $M$ is the large Majorana mass for the right-handed
neutrinos, and $m_u$ is the up-quark mass matrix), and the complete unification
of all Yukawa couplings (\ie, $\lambda_t=\lambda_b=\lambda_\tau$). (Further
model-building aspects have been extensively discussed in Ref.~\cite{NewSO10}.)
 The latter condition introduces an additional constraint
in the $(m_t,\tan\beta)$ plane, relative to the $SU(5)$ Yukawa unification case
 (see Fig.~\ref{fig:KKW}), and thus determines these two parameters, up to
the dependence on $m_b$ and $\alpha_3$ mentioned above. Typically
$\tan\beta\sim50$ is large and $m_t\sim(150-190)\GeV$ in agreement with
experiment. However, the large value of $\tan\beta$ makes it difficult to
reconcile $SO(10)$ GUTs with various phenomenological constraints \cite{SO10x}.

\subsection{Flipped SU(5)}
Finally let us discuss the ``flipped" $SU(5)\times U(1)$ model
\cite{Barr,revitalized}, where non-abelian gauge unification occurs (\ie,
$SU(2)\times SU(3)\subset SU(5)$) but part of the hypercharge $U(1)_Y$ appears
in the external $U(1)$ factor. This model is very appealing because of the many
simplifying features that it possesses over the traditional GUT models
discussed above \cite{fReviews}, and because of the special role it plays in
string model building (see Sec.~\ref{sec:strings}). Gauge symmetry breaking
down to the Standard Model gauge group occurs via vacuum expectation values of
the $H$ (\r{10}) and $\bar H$ (\rb{10}) Higgs representations. This is possible
because of the ``flipping" $u\lra d$, $u^c\lra d^c$, $e\lra\nu$, $e^c\lra\nu^c$
involved in the assignment of the Standard Model particles to the $\bar
f=\{u^c,L\}$ (\rb{5}) and $F=\{Q,d^c,\nu^c\}$ (\r{10}) representations. Thus,
$H$ and $\bar H$ contain one pair of neutral fields $\nu^c_H,\nu^c_{\bar H}$,
which get vevs along the flat direction $\vev{\nu^c_H}=\vev{\nu^c_{\bar H}}$.
The missing-partner mechanism, which can be rather cumbersome in traditional
GUTs, is here effected by the couplings $HHh$ [(\r{10})(\r{10})(\r{5})] and
$\bar H\bar H\bar h$
[(\rb{10})(\rb{10})(\rb{5})]. The resulting Higgs triplet matrix
\cite{revitalized}
\begin{equation}
\bordermatrix{
&\bar h_3&d^c_H\cr
h_3&0&\lambda_4\,\vev{\nu^c_H}\cr
d^c_{\bar H}&\lambda_5\,\vev{\nu^c_{\bar H}}&0}\ ,
\label{eq:f2/3}
\end{equation}
does not need a large non-zero (22) entry because the additional components of
the $H$ and $\bar H$ representations are eaten by the GUT gauge bosons to
become massive or become GUT Higgsinos. This natural zero mass term for $d^c_H
d^c_{\bar H}$ implies that the dimension-five proton decay operators are
negligible. Regarding see-saw neutrino masses, the right-handed neutrinos are
contained in the $F$ (\r{10}) representations.  The coupling $\lambda_1 F\bar
f\,\bar h$ provides the up-quark masses and Dirac neutrino masses, and there
are two possible sources of right-handed neutrino masses, leading to a neutrino
mass hierarchy $m_{\nu_{e,\mu,\tau}}\approx m^2_{u,c,t}/M_U$ \cite{chorus}. The
right-handed neutrino is also used to great advantage in the generation of the
baryon asymmetry of the Universe \cite{ENO}. A recent development
\cite{lowering} concerns the running of the gauge couplings and the prediction
for $\alpha_s(M_Z)$, which is naturally lowered compared to the minimal SU(5)
case and falls within the experimentally allowed range.

\section{Supergravity}
\label{sec:sugra}
As remarked in the Introduction, the much heralded convergence of the Standard
Model gauge couplings at very high energies in the presence of low-energy
supersymmetry makes sense only in the presence of a larger symmetry at
the GUT scale (at least in the field theoretical approach), {\em and} when the
supersymmetric particle masses are in the $100\GeV-1\TeV$ range. It is
remarkable that this range is consistent with present experimental lower bounds
and with the naturalness upper bounds on sparticle masses. Taken for granted in
this success of SUSY GUTs is that such sparticle masses may be obtainable in
theories of supersymmetry breaking. As is well known, models with global
supersymmetry lead to phenomenologically disastrous predictions, whereas local
supersymmetry  cures all these maladies nicely \cite{OldReviews}. Local
supersymmetry also entails the automatic incorporation of gravitational
interactions, with the spin-3/2 {\em gravitino} field as the gauge field of
local supersymmetric interactions. In analogy with the spontaneous breaking
of gauge symmetries, the massless {\em goldstino} field gets eaten by
the gravitino to acquire a mass ($m_{3/2}$). The presence of such a mass
(or order) parameter signals the breaking of supersymmetry and sets the
scale for the mass splittings of the supermultiplet partners. It is important
to realize that supergravity, even though it incorporates gravitational
interactions, it is nonetheless only an effective theory valid at scales
below the Planck mass \cite{models}. Such an effective field theory cannot
handle quantum corrections. These are parametrized by an infinite number of
non-renormalizable operators suppressed by powers of $(Q/M_{Pl})$ and endowed
with arbitrary numerical coefficients.

\subsection{Basic functions}
The rather complicated Lagrangian of supergravity theories can be described in
terms of two functions: the {\em K\"ahler function} and the {\em gauge kinetic
function}. The K\"ahler function,
\begin{equation}
G=K+\ln|W|^2\ ,
\label{eq:G}
\end{equation}
depends in turn on the {\em K\"ahler potential} ($K$) and the usual
superpotential ($W$). The K\"ahler potential is a function of all the matter
fields and their complex conjugates (\eg, $K=\phi\phi^\dagger$), whereas
the superpotential may depend on only a subset of the fields which must all
have the same chirality (\eg, $W\ni\phi\phi\phi$; while $\phi\phi\phi^\dagger$
is not allowed). For phenomenological purposes, the K\"ahler function enters in
the calculation of the scalar potential, in the normalization of the fields,
and in the calculation of the gravitino mass. The scalar potential is given by
\begin{equation}
V=e^G\left(G^IG_I-3\right)+|D|^2\ ,
\label{eq:V}
\end{equation}
where the sum (over $I$) runs over all scalar fields in the theory,
$G_I=\partial_I G$, $G^I=G^{I\bar J}G_{\bar J}$, and $G^{I\bar J}$ is the
inverse of $G_{I\bar J}$. The ``D-terms" ($|D|^2$) are analogous to those in
the case of global supersymmetry in Eq.~(\ref{eq:Vsusy}), whereas the
``F-terms" are more complicated now. The kinetic term for the scalar fields,
$G_{I\bar J}\partial^\mu\phi_I\partial^\mu\phi^\dagger_{\bar J}$ depends on
$G_{I\bar J}=K_{I\bar J}$, and determines the proper normalization of the
fields. The gravitino mass is given by
\begin{equation}
m^2_{3/2}=e^{\vev{G}}=e^{\vev{K}}\,\vev{|W|}^2\ ,
\label{eq:m3/2}
\end{equation}
in units of the reduced Planck mass $M=M_{Pl}/\sqrt{8\pi}$. Inserting
the desired form for $G$ into Eq.~(\ref{eq:V}) one can compute the scalar
masses (\eg, the masses of the squarks, sleptons, etc.). The simple choice
$K=\sum_i\phi_i\phi^\dagger_i$ gives the same (universal) mass ($m_0=m_{3/2}$)
to all scalar fields.

Derivatives of the gauge kinetic function ($f_{ab}$) determine the gaugino
masses, which in the simplest models are also universal ($m_{1/2}$).
Supersymmetry breaking manifests itself also in the all-scalar interactions,
which are patterned after the Yukawa interactions in the superpotential,
although each interaction term is accompanied by a new scalar coupling
coefficient. The universality assumption entails that all the interactions of
the same dimension possess the same supersymmetry breaking coefficient: the
trilinear scalar coupling $A_0$, and the bilinear scalar coupling $B_0$. A
supergravity theory with universal supersymmetry-breaking terms is then
described in terms of four parameters
\begin{equation}
m_0,m_{1/2},A_0,B_0\ .
\label{eq:susyx}
\end{equation}
In addition one has the parameters in the superpotential: the fermion Yukawa
couplings and the Higgs mixing term $\mu$, which do not break supersymmetry.
Specific examples of the functions $G$ and $f$ will be given in
Sec.~\ref{sec:strings} when discussing the physics of superstring models, where
these functions can be calculated from first principles, and where the simple
result in Eq.~(\ref{eq:susyx}) may not hold.

\subsection{No-scale supergravity}
An important unresolved problem in physics is the issue of the cosmological
constant ($\Lambda$) or vacuum energy ($\Lambda^4$) \cite{Weinberg}.
Observationally, the vacuum energy is seen to be extremely small (at least in
the present cosmological epoch): $\Lambda\lsim10^{-31}M_{Pl}\sim10^{-3}\eV$.
This rather small mass scale is unusual in particle physics, although light
neutrino masses obtained from the see-saw mechanism tend to reproduce such
values. Whether the cosmological constant is exactly zero or extremely small
is a basic question which particle physics and cosmology have been facing ever
since Einstein first introduced it. It is reasonable to expect that such a
small number is not the result of an incredible fine tuning, but rather the
result of exact or approximate symmetries. This line of thought would then
demand that the vacuum energy be vanishingly small at all times, and in
particular at the very high mass scales involved in supergravity
models.\footnote{A dynamical alternative to this static scenario is to assume
that the cosmological constant varies with time \cite{OldLambda}. In the long
time elapsed since the Big Bang, $\Lambda$ would have managed to reduce itself
to the very small values of interest today \cite{ESM}. This scenario seems
favored by non-critical string theory considerations \cite{ABEN,EMN,LN}.}

\begin{figure}[t]
\vspace{4.5in}
\includegraphics{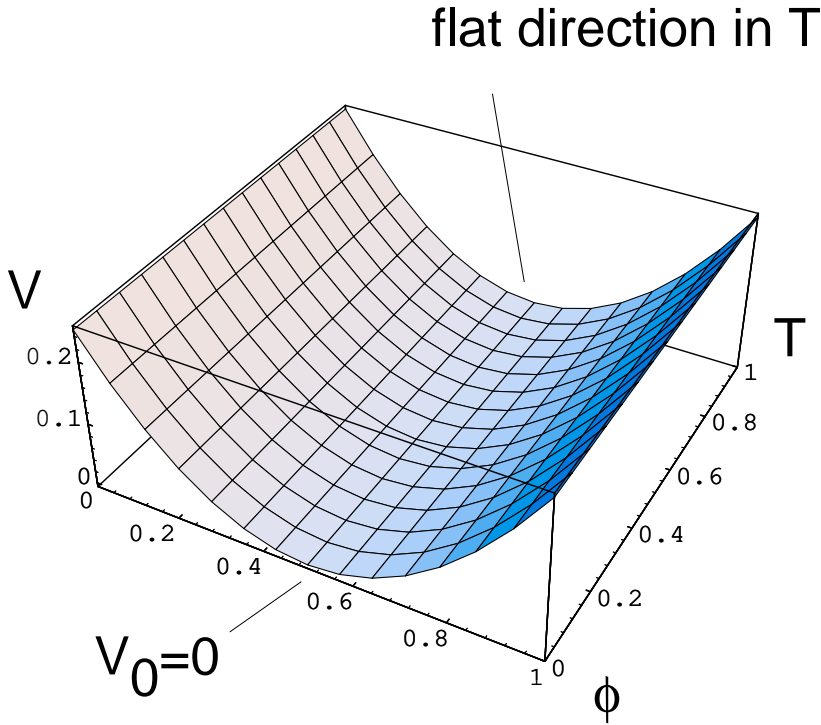}
\caption{An example of a scalar potential of the no-scale supergravity
type, where the vacuum energy vanishes ($V_0=0$) along a flat direction.}
\label{fig:flat}
\end{figure}

There is a particular class of supergravity models where the vacuum energy
$V_0=\vev{V}$ actually vanishes (at least at tree-level) \cite{Cremmer}. In
these {\em no-scale supergravity} models \cite{Lahanas,EKNI+II,LNreview}, a
judicious choice of K\"ahler potential ($K$) accomplishes this feat. The
K\"ahler potential must depend on all fields in the theory, since otherwise
their normalizations would be ill-defined. However, most fields have zero
vacuum expectation values (vevs) and for purposes of calculating the vacuum
energy they can be neglected. Moreover, it is customary to think of
supergravity theories as having an {\em observable sector} and a {\em hidden
sector}, with the two communicating with each other only through gravitational
interactions. As we discuss shortly, the hidden sector is the one usually
assumed to include the physics of supersymmetry breaking, which would happen
dynamically when certain fields gain vacuum expectation values. In the simplest
no-scale supergravity models the hidden sector is assumed to consist of a
single field $T$, which has no superpotential interactions and has a K\"ahler
potential
\begin{equation}
K=-3\ln(T+\bar T)\ .
\label{eq:Knsc}
\end{equation}
{}From Eq.~(\ref{eq:V}) it then follows that $G^TG_T\equiv 3$ for all values of
$T$, and thus at the minimum (where all other fields have zero vevs) $V_0=0$.
Furthermore, the value of $\vev{T}$ is undetermined, \ie, we
have a flat direction. This behavior is illustrated in Fig.~\ref{fig:flat}.
One can also show that there is a (modular) symmetry under which the K\"ahler
potential remains unchanged: $SU(1,1)/U(1)$ in this case. This symmetry is
important in generalizations to more complicated no-scale models
\cite{nsc,LNY94,FKZ}, but it does not guarantee the vanishing of the vacuum
energy, which depends crucially on the explicit number `3' in Eqs.~(\ref{eq:V})
and (\ref{eq:Knsc}). Note that the gravitino mass
$m^2_{3/2}=\vev{W}^2/\vev{T+\bar T}^3$, and thus the scale of supersymmetry
breaking, is also undetermined. This property can be exploited to implemented
the {\em no-scale mechanism} \cite{Lahanas,Zwirner}, whereby physics at the
electroweak scale ``bends" this flat direction and determines the preferred
value of  $\vev{T}$ and therefore of $m_{3/2}$. It is important to realize that
for the no-scale mechanism to preserve the desired hierarchy ($m_{3/2}\ll
M_{Pl}$), there should not appear terms in the scalar potential with
dimensional coefficients containing large mass scales \cite{EKNI+II,FKZ,LNnsc}.
This is satisfied automatically at the tree level. At the one-loop level there
appears a term $\propto{\rm Str}\,{\cal M}^2\,
\Lambda^2$, with $\Lambda\sim M_{Pl}$. A consistency requirement is then that
the spectrum satisfy the sum rule ${\rm Str}\,{\cal M}^2=0$, after
supersymmetry breaking \cite{EKNI+II,FKZ}. Further consistency constraints
at higher loops have been considered also \cite{Bagger}.

\subsection{Supersymmetry breaking}
Above we have seen how supersymmetry breaking is parametrized in terms of
the gravitino mass, with specific spectra obtained for given choices of
$G$ and $f$. Equation (\ref{eq:m3/2}) shows that in order to have supersymmetry
breaking (\ie, $m_{3/2}\not=0$), one must have $e^{\vev{K}}\not=0$, which
is satisfied in all cases of interest, and $\vev{W}\not=0$. The latter then
becomes the real pre-requisite for supersymmetry breaking. In fact,
non-renormalization theorems show that if supersymmetry is not broken at the
tree level, then it is not broken at any order in perturbation theory.
In the latter case supersymmetry could still be broken by non-perturbative
effects. Both approaches have been pursued in the literature.

Tree-level breaking is accomplished in certain superstring models
\cite{Coordinate}, where $\vev{W}=c$, with $c$ some ${\cal O}(1)$ constant. In
this case our simple
no-scale model in Eq.~(\ref{eq:Knsc}) gives
\begin{equation}
m^{\rm tree}_{3/2}\sim {c\over(\kappa T+\kappa\bar T)^{3/2}}\,M_{Pl}\ ,
\label{eq:m3/2tree}
\end{equation}
where $\kappa\sim1/M_{Pl}$. Since $c\sim1$, $\kappa T$ must be large to produce
a sufficiently small value of $m_{3/2}$. One can show that this is equivalent
to a decompactification limit, where some of the superstring internal
dimensions become nearly macroscopic \cite{ABLT}. Such scenario would have
manifold observable consequences at the next generation of high-energy particle
accelerators \cite{ABQ}.

A more `popular' approach to supersymmetry breaking entails the inclusion of
non-perturbative contributions in $W$. This scenario has been extensively
explored mostly in the context of strongly-interacting (hidden sector) gauge
field theories \cite{GaugCond,Quevedo}, as opposed to the little understood
string non-perturbative interactions. In this scenario one assumes that the
hidden sector is composed of a gauge theory with or without matter particles.
This sector of the theory is chosen such that as the mass scale is lowered, the
gauge coupling increases (like in the usual QCD). The scale at which the gauge
coupling blows up (using the one-loop approximation to the beta function
$\beta$) is called the {\em condensation scale}
\begin{equation}
\Lambda=M e^{-8\pi^2/|\beta| g^2}\ ,
\label{eq:Lambda}
\end{equation}
where the gauge coupling takes the value $g$ at the string or Planck scale
$M$. At the condensation scale the strongly-interacting dynamics typically lead
to the condensation of gauginos ($\vev{\lambda\lambda}\not=0$) and, if light
matter is also present, to the formation of ``meson"
($\vev{H\bar H}$) bound states. Below the condensation scale, the theory
described in terms of these objects generates a non-perturbative superpotential
with non-zero vev $\vev{W_{\rm np}}\sim \Lambda^3$, which breaks supersymmetry
if certain conditions are met \cite{Quevedo}. Moreover, the gravitino mass
(see Eq.~(\ref{eq:m3/2})) is then exponentially suppressed relative to the
Planck scale
\begin{equation}
m^{\rm np}_{3/2}\sim {e^{-24\pi^2/|\beta|g^2}
\over(\kappa T+\kappa\bar T)^{3/2}}\,M_{Pl}\ ,
\label{eq:m3/2np}
\end{equation}
and of the desired magnitude for the `natural' values of $\kappa T\sim1$
(\ie, those preferred by $T$-duality considerations).
The resulting scalar potential may or may not generate a large vacuum energy,
depending on the model details. It may also not lift the $T$-flat direction,
which may be determined by the usual no-scale mechanism.

The supergravity models described above are quite general, and thus rather
unpredictive. One lacks the ability to calculate from first principles the
K\"ahler potential, the superpotential, and the gauge kinetic function,
although simple assumptions seem to work  well.  To proceed we must
resort to a theory where gravity can be consistently quantized. The only
known example is string theory. String theory has further advantages, one
of which is the ability to calculate the above unknown functions. At the same
time, string supergravity is complicated by the appearance of {\em moduli}
fields and new symmetries, which do away with the simple ans\"atze made in
traditional supergravity.

\section{Superstrings}
\label{sec:strings}
The path that we have followed so far --  in the direction of ever increasing
mass scales -- has not yet provided a means of calculating the many unknown
parameters of the Standard Model or its supersymmetric extensions, even
though a great deal of synthesis has been accomplished at the various stages,
implying that fewer parameters are required as we uncover larger and larger
symmetries. This path has also led to the inclusion of gravitational
interactions in the picture of elementary particle physics, although only
at an effective level below the Planck mass. Superstring theory \cite{GSW,
Formulations} is usually described as the only known theory where quantum
gravitational corrections can be consistently computed. This is accomplished by
a drastic change in our picture of the particle world, which is now viewed as
consisting of a single type\footnote{There are in fact a few types of strings
one can consider (\eg, Type I, Type II, heterotic), although recent
understanding of superstring dualities appears to indicate that all these are
different manifestations of the same underlying ``M-theory" \cite{Schwarz}.} of
Planck-sized string ($\ell_{Pl}\sim10^{-33}\cm$), with the various particles
represented by different modes of string vibration. The finite size of the
string provides a cut-off in the distance scale (it cannot be arbitrarily
small), which softens the ultraviolet divergences of conventional quantum field
theory. String perturbation theory is envisioned as an expansion on the
topology (genus) of the two-dimensional world-sheet that describes string
propagation: a (no-hole) sphere at tree level, a (one-hole) torus at one loop,
a two-hole surface at two loops, etc. This topological expansion is much
simplified compared with field theory, as in string theory there is a single
``Feynman diagram" at each order in perturbation theory.\footnote{Such
simplifying properties of string perturbation theory have been successfully
exploited to compute complicated many-particle diagrams in field theory
\cite{BK}.} One could not ask for a more unified theory: not only are the
interactions unified, but so are the particles themselves. This theory
possesses only one parameter: the Planck mass; the various unknown low-energy
parameters must be calculable from first principles.

\subsection{String model building}
In practice string theory is only partially understood. This state-of-affairs
is most evident in the very large number of solutions to the string equations
(vacua or ``string models") that are known to exist. With our limited
understanding, all these solutions appear equally acceptable from the
theoretical point of view. In contrast, it is found that phenomenology
is very discriminating, basically wiping out all known solutions, although
some fare much better than others. String model-builders are then charged with
finding the best possible string model. This task was started in 1984, when
the first consistent string solutions were found by Green and Schwarz with
the gauge group SO(32) in ten dimensions \cite{GS}. Shortly thereafter the
heterotic string was introduced \cite{heterotic}. Most of the work since then
has concentrated on exploring different {\em compactification} schemes to
reduce the theory from ten dimensions down to four. Within a fixed
compactification scheme (Calabi-Yau manifolds \cite{CY}, orbifolds
\cite{SymOrb,AsymOrb}, free-fermionic constructions \cite{FFF}, etc.) one can
build consistent models and study their phenomenological properties.

A common feature in string model-building is the need to provide as inputs
parameters describing the two-dimensional world-sheet, which underlies the
four-dimensional world according to string theory. This indirect input method
makes string model-building less ``intuitive" than conventional GUT model
building, and harder because of the severe consistency constraints that need to
be satisfied. In any given compactification scheme, after the two-dimensional
inputs are provided, well defined procedures can be applied to obtain the
four-dimensional results. In the case of free-fermionic strings, the procedure
involves a large amount of rather simple algebraic manipulations, making it
amenable to automation. In a typical situation, the resulting model consists of
a gauge group with several gauge factors, and set of massless (and massive)
matter representations, such that all anomalies automatically cancel. Each
state in the model can be represented by a {\em vertex operator} which
encompasses all the gauge degrees of freedom, as well as important
two-dimensional quantum numbers which appear as global quantum numbers from the
four-dimensional point of view. These ``hidden" quantum numbers restrict the
possible interactions among the fields beyond the usual gauge symmetry
constraints, leading to otherwise unexplainable zero couplings (or
``textures"). In this way it is possible to calculate explicitly the
contributions to the superpotential (at cubic and higher orders) and the
K\"ahler potential, in any given model. A subtlety in this process arises by
the presence of special fields called moduli, which have no scalar potential --
they parametrize flat directions. It is important to identify these fields
because the K\"ahler potential can then be recast (through field redefinitions)
in a more useful form, which makes manifest the presence of the moduli and
their corresponding modular symmetries (see \eg, \cite{LNY94}).

One of the more basic model-building choices to be made is the {\em level} of
the Kac-Moody algebra of two-dimensional currents that underlies the
four-dimensional gauge group \cite{GO}. This level ($k$) is a positive integer,
which for most of the history of string model-building was chosen implicitly or
explicitly to be unity ($k=1$). More complicated constructs are required to
build models with levels greater than one \cite{Lewellen}. The choice of level
of the gauge group has a dramatic phenomenological implication: the smaller the
level, the smaller the set of allowed massless representations in a possible
consistent model. This is a general property of string models, and becomes most
restrictive at level one ($k=1$) allowing only \cite{FIQ,ELN} SO(2n): singlet,
vector, and spinor representations; SU(n): totally antisymmetric
representations, as shown in Table~\ref{Table2}; $E_6$: \r{1},\r{27},\rb{27};
$E_7$: \r{1},\r{56}; $E_8$: \r{1}. Note that the traditional GUT-breaking
(adjoint) Higgs representations are not allowed at level one. They become
allowed at level two or higher. In fact, it has been recently become topical to
investigate methods by which level-two (or higher) models can be built. These
methods have been developed in the context of free-fermionic models
\cite{Lykken}, and symmmetric \cite{Aldazabal} and asymmetric \cite{KT}
orbifold constructions. As mentioned above, methods to build level-one models
are manifold \cite{CY,SymOrb,AsymOrb,FFF} and have been known for some time.

\begin{table}[t]
\caption{Allowed massless representation in SU(n) gauge groups realized
with level one Kac-Moody algebras.}
\label{Table2}
\bigskip
\hrule
\begin{center}
\begin{tabular}{l|l}
$n$&Representation\\ \hline
2&\r{1},\r{2}\\
3&\r{1},\r{3},\rb{3}\\
4&\r{1},\r{4},\rb{4},\r{6}\\
5&\r{1},\r{5},\rb{5},\r{10},\rb{10}\\
6&\r{1},\r{6},\rb{6},\r{15},\rb{15},\r{20}\\
7&\r{1},\r{7},\rb{7},\r{21},\r{21},\r{35},\rb{35}\\
8&\r{1},\r{8},\rb{8},\r{28},\rb{28},\r{56},\rb{56},\r{70}\\
9&\r{1},\r{9},\rb{9},\r{36},\rb{36},\r{84},\rb{84}\\
10--23&\r{1},\r{n},\rb{n},{\bf n(n--1)/2},\rb{n(n-1)/2}
\end{tabular}
\end{center}
\bigskip
\hrule
\end{table}

\subsection{String unification}
{\tt To GUT or not to GUT}? \cite{John} The traditional motivation for GUTs,
\ie, their prediction of the unification of all gauge couplings, turns out to
be automatic in string models (up to factors of the level of the respective
gauge groups) \cite{Ginsparg}. Thus, it is not obvious that a string-derived
GUT is particularly compelling. In fact, such models require higher-level
constructions, which so far have met with limited phenomenological success in
the areas of the number of generations and the doublet-triplet splitting
problem \cite{Lykken,Aldazabal}. In any event, string unification (to lowest
order) is predicted to occur at the scale $M_{\rm string}\approx 5\times
g\times 10^{17}\GeV$ \cite{Kaplunovsky}, where $g$ is the unified gauge
coupling. Above this scale the spectrum of massive string particles is excited
and the conventional field theory description fails. Nonetheless, it is
possible to calculate the ``threshold" effects of these particles
\cite{Kaplunovsky,DKL,AGNT,DF}, which entail splittings among the various gauge
couplings at $M_{\rm string}$, or equivalently, a shift in the effective
unification scale.

An important question in string model-building is how to reconcile the string
unification scale ($M_{\rm string}$) with the simplest SUSY GUTs unification
scale ($M_{\rm LEP}\sim10^{16}\GeV$), which is some twenty times smaller. Such
``discrepancy" may disappear once string models are better understood, although
in the meantime a few solutions to bridge this gap have been proposed, such as
adding new intermediate-scale ``gap" particles \cite{price} or allowing the
string threshold corrections to decrease the effective string unification scale
down to $M_{\rm LEP}$ \cite{Ibanez}. The latter scenario appears now
disfavored, as it requires large values of the moduli fields that parametrize
the threshold corrections, which are hard to obtain in actual string models
\cite{moduli,DF}, and still requires the addition of new particles beyond the
MSSM \cite{DF}. A recent proposal in the context of flipped SU(5) takes
advantage of several stringy features of the model and yields a natural
scenario for string unification, along the lines of the ``gap" particle
scenario \cite{TwoStep}.

\subsection{Dilaton and S-duality}
A manisfestation of the ``no-parameter" character of string models is the
value of the gauge coupling, which is determined dynamically by the vacuum
expectation value of the dilaton field $S$: $g^2=1/{\rm Re}\vev{S}$, with
$\vev{S}$ in Planck units. The dilaton is a modulus field, which has no
potential at any order in perturbation theory: the gauge coupling slides along
this flat direction. A nagging question in string theory is how to determine
$\vev{S}$. In the mechanism of supersymmetry breaking via gaugino condensation,
field theory non-perturbative effects involve $S$, but typically along a
runaway direction (\ie, $\vev{S}\to\infty$). This problem may be solved by
tuning two gaugino condensates such that their competing effects stabilize
$\vev{S}$ \cite{Krasnikov}. In practice, such models have proved difficult to
construct in string model building. On the other hand, string non-perturbative
effects are expected to play a major role in the determination of $\vev{S}$.
The most significant progress on this question has come from the assumption
that the dilaton obeys duality symmetries similar to those obeyed by the
traditional moduli fields \cite{FILQ}. This ``$S$-duality" entails specific
forms for the $S$-dependence of the scalar potential, and typically predicts
$\vev{S}\sim1$: a very desirable result. This symmetry has far-reaching
consequences, as it entails transformations such as $S\to 1/S$, which connect
the weakly-interacting to the strongly-interacting regimes of string theory.
Recent work in this direction has led to the discovery of dualities (of $S$ and
$T$ types) connecting strings to higher-dimensional objects called membranes,
and to dualities among different kinds of strings (\eg, heterotic and Type II).
This topic has become very active recently \cite{Schwarz} and is likely to
greatly illuminate our understanding of string theory, especially in its
non-perturbative regime. For our purposes, we hope that the ultimate picture
that emerges will still allow for a meaningful perturbative approach to string
model building.

\subsection{Realistic models}
String models have been built using several different string formulations
\cite{Formulations}. Originally Calabi-Yau compactification \cite{CY}
 was the preferred construction, resulting in models with gauge groups such as
$SU(3)^3$ \cite{SU3}. Later symmetric \cite{SymOrb} and asymmetric
\cite{AsymOrb} orbifold constructions were found to be more mathematically
accessible, and models with the Standard Model gauge group were constructed
\cite{SMOrb}. A sizeable fraction of the string model-building effort
has been carried out in yet another construction: the free-fermionic
formulation \cite{FFF}, where models with the gauge groups SU(5)$\times$U(1)
\cite{revamp,search}, SU(4)$\times$SU(2)$\times$SU(2) \cite{ALR}, and the
Standard Model \cite{SMlike} gauge group have been constructed. A large amount
of effort has been devoted to the study of these models, where the
superpotential has been determined at the cubic and non-renormalizable levels
\cite{KLN}, and the K\"ahler potential has become available recently
\cite{LNY94}. There is no room here to discuss the properties of these models
in any detail. However, a few important properties can be mentioned, such as
their level-one nature, which implies that no adjoint representations are
required to break their unified gauge groups. One also has the unparalleled
ability to calculate the couplings in the superpotential, in particular the
fermion Yukawa couplings. A typical prediction is $\lambda\sim g\sim1$,
which implies a quark mass in the range $m_q\sim (150-200)\GeV$
\cite{revamp,SMlike,Zero}. Such prediction agrees with experiment for the top
quark, and thus one should {\tt ask not why the top quark is so heavy}, but
instead {\tt ask why the other quarks are so light}. The remaining quarks may
have suppressed Yukawa couplings, principally because of several stringy
selection rules stemming from the ``hidden" quantum numbers discussed above.
These couplings would vanish at the cubic level but would arise at higher
orders in superpotential interactions, suppressed by powers of $M_{\rm
string}/M\sim{1\over10}$. This desirable ratio \cite{revamp}
is generated in the presence of a seemingly anomalous $\rm U_A(1)$ factor in
the gauge group, which forces the theory into a nearby vacuum where some scalar
fields gain vacuum expectation values \cite{DSW}. This mechanism to generate
a hierarchical fermion mass spectrum has inspired recent attempts at
constructing textured fermion mass matrices \cite{textures}.

\subsection{String supergravity}
With the knowledge gained from strings, low-energy effective theories can
be constructed in the form of standard supergravity theories, but with
calculable forms for the K\"ahler potential, superpotential, and gauge kinetic
function \cite{KWf,DKL}. This exercise has turned out to be more subtle than
naively expected because of the duality symmetries that string models possess
to all orders in perturbation theory. These symmetries are not so evident at
lowest order in perturbation theory, but one can invoke general arguments and
rewrite the tree-level results so that the duality symmetry is manifest. For
instance, the lowest order form for the K\"ahler potential in a typical model
is
\begin{equation}
K=\phi\phi^\dagger+\coeff{1}{2}\phi\phi^\dagger\phi\phi^\dagger+\cdots
=-\ln(1-\phi\phi^\dagger)\ .
\label{eq:KLO}
\end{equation}
Direct calculation yields the first two terms in this expression, whereas the
logarithm is the presumed all-orders result obtained from duality symmetry
considerations. Duality symmetries also arise in the calculation of the
superpotential, especially at the non-renormalizable level (see \eg,
\cite{modinv}). A more dramatic result is obtained in the case of the gauge
kinetic function $f$, which receives a universal tree-level contribution of the
form $f=kS$ ($k$ is the level of the Kac-Moody algebra), whereas considerations
of duality anomalies show that it receives readily-calculable one-loop
corrections only \cite{f1loop}. Duality symmetries also have a ``down" side, in
that one needs to understand how they are broken, \ie, what is the expectation
value of the moduli fields, as otherwise every observable remains undetermined.

I conclude this section by discussing a particular class of string models,
those that respect the postulates of no-scale supergravity:
(i) the (tree-level) vacuum energy vanishes, (ii) there is a flat direction
along which the gravitino mass is undetermined, and (iii) the scalar potential
does not depend quadratically on large mass scales (\ie, ${\rm Str}\,{\cal
M}^2=0$). Traditional supergravity models with these properties have been
discussed in Sec.~\ref{sec:sugra}, whereas there has been recent progress in
studying string models with these properties \cite{nsc,FKZ,LNnsc}. (In fact,
the no-scale supergravity structure was identified early on as a generic
property of string supergravities \cite{Witten}.) In string models these
constraints are quite restrictive: the (tree-level) K\"ahler potential takes
the form $K=-\ln(S+S^\dagger)-2\ln(T+T^\dagger)$, whereas the spectrum of the
model needs to be correlated with the corresponding gauge group in a special
way, if the third constraint (${\rm Str}\,{\cal M}^2=0$) is to be satisfied
\cite{LNnsc}. The problem becomes more subtle when one considers realistic
models with anomalous $\rm U_A(1)$ factors in the gauge group, in which case it
has been possible to construct the first semi-realistic string models where the
third postulate is satisfied \cite{StrM^2}. Given the large number of string
models, it appears sensible to apply reasonable constraints to reduce the
number of possible realistic models. String no-scale supergravity is an
interesting example of such endeavor.

The above discussion has focused on {\em critical} string theory, implicitly
assuming a flat gravitational background. This need not be the case, and
certainly was not the case during the early universe. {\em Non-critical}
string theory is required to describe such situations. This subject is rather
interesting, as it introduces a dependence on the dynamical time that
parametrizes the approach to the flat background (\eg, the ``cosmic" time
elapsed since the Big Bang). A variety of possible observable consequences have
been studied in a class of such models \cite{ABEN,EMN}, such as generic
violation of CPT, the collapse of the wavefunction in quantum mechanics, the
time-dependence of the fundamental constants, a new model of inflation, etc.
\cite{Time}.

\section{Dynamics}
\label{sec:dynamics}
Being able to construct supersymmetric models of particle physics at very
high energies is the first step in making contact with experimental reality.
One must also take into account the fact that experiments are performed at
energies ($\sim1\TeV$) much lower than those at which the models are most
naturally built ($\sim10^{16-18}\GeV$). This means that the model parameters
need to be ``evolved" down  through a large ratio of scales:
$10^{16-18}/10^2\sim10^{14-16}$. The underlying quantum field theory, upon
which our gauge theories are built, provides a precise prescription for such
dynamical evolution through the use of the {\em renormalization group
equations} (RGEs). These equations encode the scale dependence of the model
parameters, which is necessary to maintain the renormalization group invariance
of the theory as a whole. All model parameters (gauge and Yukawa couplings,
scalar masses and couplings, etc.) participate in this set of coupled
first-order linear differential equations, with one equation per parameter. The
coefficients in these equations can be calculated order-by-order in
perturbation theory, although in practice are known only to one- or two-loop
order \cite{RGEs}. This renormalization-group scaling takes into account the
largest contributions to such evolution, those coming from the large logarithms
$\ln(M_U/M_Z)$. The best known evolution equations describe the running of the
gauge couplings, as discussed in Sec.~\ref{sec:GUTs}.

\subsection{Radiative electroweak symmetry breaking}
In the process of RG evolution to lower energies one may encounter two
phenomena: decoupling of particles and gauge symmetry breaking. When the
running scale falls below the mass of a given particle, such particle is
dropped from the subsequent evolution by means of some decoupling procedure.
This procedure must be followed repeatedly in the $100\GeV$--$1\TeV$ range,
where most of the supersymmetric particles decouple. A more drastic procedure
must be followed when the breaking of a gauge symmetry is encountered,
typically the electroweak symmetry at $\sim100\GeV$. In fact, our dynamical
picture would be incomplete if as we lower the running scale we did not observe
signs that the electroweak symmetry is broken, \ie, that the Higgs mechanism is
happening. In the context of supersymmetric unified theories, the Higgs
mechanism occurs dynamically as the appropriate mass parameters in the
supersymmetric Standard Model scalar potential evolve with scale, and
eventually change sign near the electroweak scale. This {\em radiative
electroweak breaking} phenomenon \cite{EWx} depends crucially on supersymmetry,
supersymmetry breaking, and the running of mass parameters down from a large
mass scale.

\begin{figure}[t]
\vspace{4in}
\includegraphics{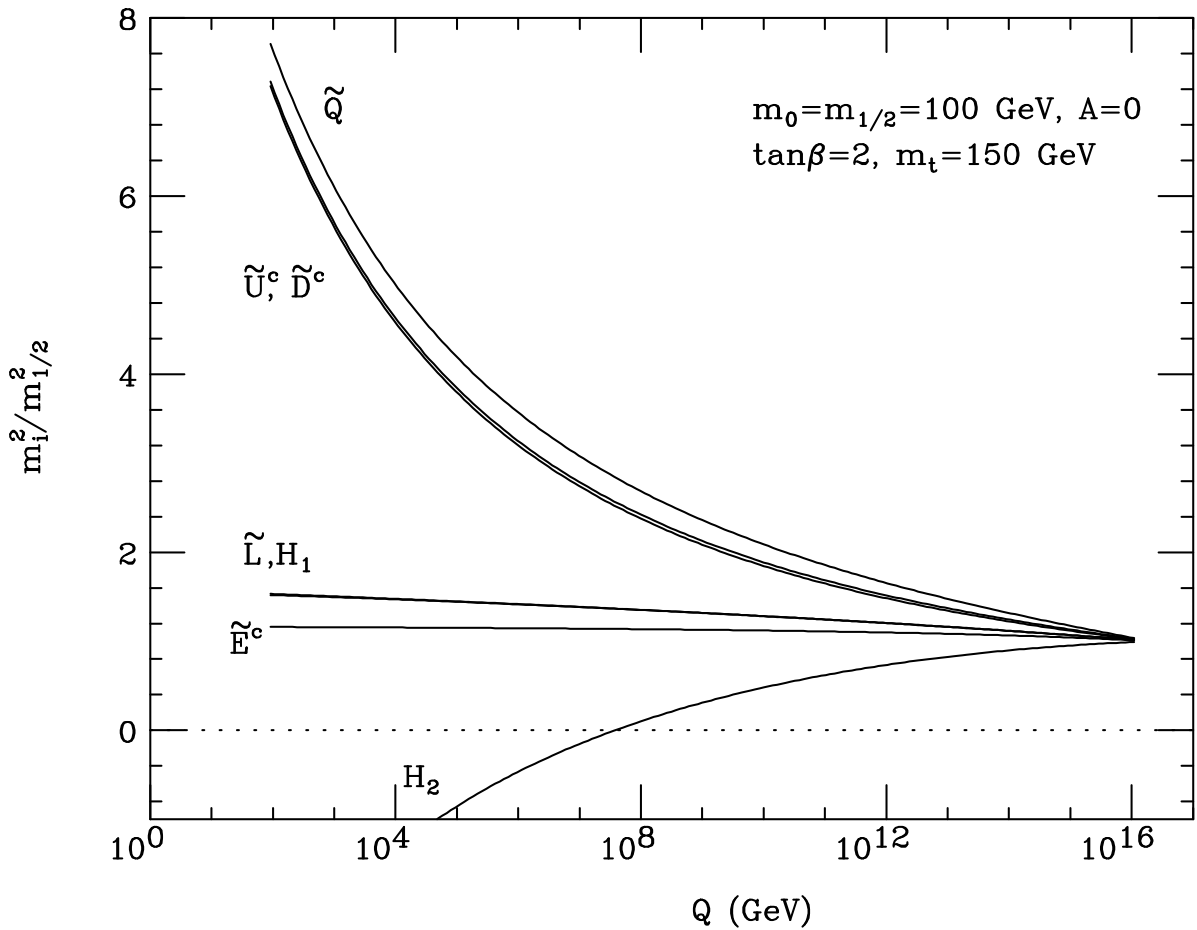}
\vspace{1in}
\caption{Running of the scalar masses in supergravity for typical values of
the parameters. Note that for $Q<Q_0$ the electroweak symmetry is broken
($m^2_{H_2}<0$), but the $\rm SU(3)_C$ color and $\rm U(1)_{\rm em}$
electromagnetic gauge symmetries remain unbroken.}
\label{fig:masses}
\end{figure}

To illustrate the concept, let us consider a typical RGE for a scalar mass
$\widetilde m$
\begin{equation}
{d\widetilde m^2\over dt}={1\over(4\pi)^2}\left\{
-\sum_i c_i g^2_i M^2_i + c_t\lambda^2_t\left(\sum_i\widetilde
m^2_i\right)\right\}\ ,
\label{eq:RGE}
\end{equation}
where $M_i$ are the gaugino masses, and the $c$ coefficients are given below
for the various MSSM particles
\begin{equation}
\begin{array}{cccc}
&c_t&c_3&c_2\\
H_1&0&0&6\\
H_2&6&0&6\\
\widetilde Q&0&\coeff{32}{3}&6\\
\widetilde U^c&0&\coeff{32}{3}&0\\
\widetilde D^c&0&\coeff{32}{3}&0\\
\widetilde L&0&0&6\\
\widetilde E^c&0&0&0
\end{array}
\label{eq:coeffs}
\end{equation}
The result of running these RGEs is illustrated in Fig.~\ref{fig:masses}
for the indicated values of the parameters. For $Q<Q_0$, $m^2_{H_2}<0$ whereas
$m^2_{H_1}>0$. The sign change signals the breaking of the electroweak
symmetry. Note that the top-quark Yukawa coupling ($\lambda_t$) plays a  {\em
fundamental} role in driving $m^2_{H_2}$ to negative values. This is only
possible if $\lambda_t$ is large enough to counteract the effect of the gauge
couplings, and thus requires the existence of a ``heavy top quark." Note that
$m^2_{\tilde Q,\tilde U^c,\tilde D^c}$ remain positive because of the large
$\alpha_3$ contribution ($\propto c_3$) to their running. For the same reason
the sleptons ($\tilde L,\tilde E^c$) ``run" much less. Thus, this mechanism
breaks the electroweak symmetry but preserves the $\rm SU(3)_C$ color and $\rm
U(1)_{\rm em}$ electromagnetic gauge symmetries.

\subsection{Supersymmetry breaking scenarios}
In considering the low-energy predictions for the sparticle spectrum in the
context of supergravity models, several scenarios have arisen in the
literature. These scenarios correspond to specific choices of the K\"ahler
function and/or the dominant source of supersymmetry breaking. If one assumes
that the K\"ahler potential is observable-sector blind, then universal
supersymmetry-breaking mass parameters are obtained, and these are described by
the four parameters in Eq.~(\ref{eq:susyx}): $m_0,m_{1/2},A_0,B_0$. In this
context one can impose further conditions on the K\"ahler potential leading to
the special case:
\begin{equation}
m_0=A_0=0\ ,
\label{eq:nsc}
\end{equation}
sometimes referred to as ``no-scale", since the earliest no-scale supergravity
models predicted such relations \cite{EKNI+II}, although modern no-scale models
usually depart from them \cite{Zero}. If supersymmetry breaking is dominated by
the F-term of the dilaton field (a stringy effect), one obtains \cite{KL,BIM}
\begin{equation}
m_0=\coeff{1}{\sqrt{3}}\,m_{1/2}\ ,\qquad A_0=-m_{1/2}\ .
\label{eq:dilaton}
\end{equation}
One can make further assumptions concerning the origin of the Higgs mixing
parameter ($\mu$) \cite{muproblem}, and obtain predictions for $B_0$
\cite{KL,BIM}, although these are rather model dependent. In string-derived
models the K\"ahler potential has non-trivial structure, which distinguishes
between different fields via their {\em modular weights} or charges under
modular symmetries. In the simplest models of this kind, the scalar masses take
the form \cite{BIM}
\begin{equation}
m^2_i=m^2_{3/2}(1+n_i\cos^2\theta)\ ,
\label{eq:non-univ}
\end{equation}
where $n_i$ is the modular weight of the $i$-th field, and
$\tan\theta=\vev{F_S}/\vev{F_T}$ quantifies the amount of dilaton $\vev{F_S}$
and moduli $\vev{F_T}$ contributions to the supersymmetry-breaking F-term. In
the dilaton scenario of Eq.~(\ref{eq:dilaton}): $\theta\to{\pi\over2}$. The
most striking property of this result is the general lack of universality of
the scalar masses \cite{KL}. The possible choices of $n_i$ are model dependent,
although always integer. In {\em generic} orbifold models these can vary quite
a bit ($-1$ to $-5$), while in $Z_2\times Z_2$ orbifolds they are always equal
to $-1$, implying universal scalar masses automatically. In {\em explicit}
string-derived models the scalar masses can be calculated explicitly, with
results not necessarily following the simple formula in
Eq.~(\ref{eq:non-univ}). For instance, one finds models where some of the
states have common scalar masses equal to $m_{3/2}$, while the rest of the
states have vanishing scalar masses \cite{Zero}. In any event, it is clear that
the more specific models one considers, the more one seems to depart from the
naive assumption of universal scalar masses. On the other hand, these specific
models are much more predictive than the generic ones and may be easily
falsifiable.

\subsection{Mass relations}
The coupled set of renormalization group equations mentioned above must in
principle be solved numerically. However, under reasonable assumptions some
of the equations can be solved analytically.  For instance, RGEs for the
running gauge couplings in Eq.~(\ref{eq:gRGE}) can be solved exactly
to lowest order in the beta functions
\begin{equation}
{d\alpha^{-1}_i\over dt}=-{b_i\over2\pi} \quad\Rightarrow\quad
\alpha_i(Q)={\alpha_i(Q_0)\over1-{b_i\over2\pi}\alpha_i(Q_0)\ln(Q/Q_0)}\ .
\label{eq:gRGE1}
\end{equation}
In a slightly more complicated manner one can also solve the RGEs for the
first- and second-generation squark and slepton masses. In this case one
neglects the Yukawa couplings of the corresponding quarks and leptons, as
these are much smaller than those of their third generation counterparts.
One obtains
\begin{eqnarray}
\wt m_i^2=m^2_{1/2}(c_i+\xi^2_0)-d_i{\tan^2\beta-1\over\tan^2\beta+1}M^2_W
\label{eq:masses}
\end{eqnarray}
where $\xi_0=m_0/m_{1/2}$ and $d_i=(T_{3i}-Q)\tan^2\theta_w+T_{3i}$ (\eg,
$d_{\tilde u_L}={1\over2}-{1\over6}\tan^2\theta_w$, $d_{\tilde
e_R}=-\tan^2\theta_w$). The $c_i$ coefficients can be
calculated numerically in terms of the low-energy gauge couplings, and are
given in  Table \ref{Table3} for $\alpha_3(M_Z)=0.118$ and two
GUT choices: standard minimal SU(5) unification at the scale $\sim10^{16}\GeV$,
and a string-inspired unification at the scale $\sim10^{18}\GeV$. In the latter
case a minimal set of additional matter representations has been introduced to
delay unification (a vector-like quark doublet $Q,Q^c$ and a vector-like quark
singlet $D,D^c$) \cite{price}. In the table we also give $c_{\tilde
g}=m_{\tilde g}/m_{1/2}$. The above approximation fails for the third
generation sparticles, especially when $\tan\beta$ is large, since then the $b$
and $\tau$ Yukawa couplings are enhanced and can be as large as the top-quark
Yukawa coupling; analytical expressions are however still obtainable \cite{FL}.

\begin{table}[t]
\caption{Values of the $c_i$ and $d_i$ coefficients in Eq.~(35)
[$\alpha_3(M_Z)=0.118$] for the first- and second-generation sfermions (\ie,
$c_{\tilde e_R}=c_{\tilde\mu_R}$, and so on) for the minimal SU(5) GUT model
and for a string-inspired GUT (GUST) that unifies at the string scale. Also
shown is $c_{\tilde g}\equiv m_{\tilde g}/m_{1/2}$.}
\label{Table3}
\bigskip
\hrule
\begin{center}
\begin{tabular}{cccc}
$c_i$&GUT &GUST&$d_i$  \\ \hline
$c_{\tilde e_R}$&.149&.143&$-\tan^2\theta_W$\\
$c_{\tilde e_L}$&.512&.402&$-{1\over2}+{1\over2}\tan^2\theta_W$\\
$c_{\tilde \nu}$&.512&.402&${1\over2}+{1\over2}\tan^2\theta_W$\\
$c_{\tilde u_L}$&6.28&3.91&${1\over2}-{1\over6}\tan^2\theta_W$  \\
$c_{\tilde d_L}$&6.28&3.91&$-{1\over2}-{1\over6}\tan^2\theta_W$  \\
$c_{\tilde u_R}$&5.87&3.60&${2\over3}\tan^2\theta_W$  \\
$c_{\tilde d_R}$&5.82&3.55&$-{1\over3}\tan^2\theta_W$\\
$c_{\tilde g}$&2.77&2.01 &
\end{tabular}
\end{center}
\bigskip
\hrule
\end{table}

In unified supergravity models with radiative electroweak breaking \cite{REWx}
there are many predicted masses in terms of a few input parameters, entailing
several mass relations. A particularly important one concerns the masses of the
squarks of the first two generations. From Eq.~(\ref{eq:masses}) we see that
for the squark masses of current interest ($m_{\tilde q}\gsim200\GeV$), the
second (``D-term") contribution is small relative to the first one because the
corresponding $c_i$ are large (see Table~\ref{Table3}). Thus, all squark masses
are nearly degenerate and one usually talks about an average squark mass. On
the other hand, the top-squark masses are obtained by diagonalizing a
$2\times2$ matrix with off-diagonal entries proportional to the top-quark mass
(\ie, $m_t(A_t+\mu/\tan\beta)$), and thus the lightest eigenvalue ($\tilde
t_1$) can easily be much lighter than all the other squarks. In contrast, one
does not usually talk about an average slepton mass because the corresponding
$c_i$ coefficients are much smaller (see Table~\ref{Table3}), making the
sleptons typically lighter (or even much lighter) than the squarks. However,
the sleptons can be as heavy as the squarks as long as the other relevant
parameter ($\xi_0$) is large enough. This spectrum of squark and slepton masses
could in principle be measured accurately  (``sparticle spectroscopy"
\cite{FHKN}) at a suitable facility, such as the planned $e^+e^-$ next linear
collider (NLC) \cite{Peskin} or a recently proposed $\mu^+\mu^-$ collider
\cite{Bargermu-mu}.

In a unified theory one also gets a relation among the gaugino masses
\begin{equation}
{M_1\over\alpha_1}={M_2\over\alpha_2}={m_{\tilde
g}\over\alpha_3}={m_{1/2}\over\alpha_U}\ ,
\label{eq:gauginos}
\end{equation}
where $M_1,M_2$ are the U(1) and SU(2) gaugino masses, and $\alpha_U$ is
the gauge coupling at the unification scale. The first relation gives
$M_1={5\over3}\tan^2\theta_W M_2$, whereas the second one gives
$M_2=(\alpha_2/\alpha_3)m_{\tilde g}$, which is referred to by experimentalists
as the ``GUT relation". In fact, these relations allow one to connect
experiments at hadron colliders ($m_{\tilde g}$) with experiments at $e^+e^-$
colliders ($m_{\chi^0_{1,2,3,4}},m_{\chi^\pm_{1,2}}$), and has been used to
set a lower limit on the lightest neutralino mass $m_{\chi^0_1}\gsim20\GeV$
\cite{Roszkowski}.
Experimental verification or falsification of these mass relations will provide
a direct window into the physics at the GUT scale, in particular the gauge
group and the gauge kinetic function. Another kind of mass relation which
follows in supergravity models concerns the chargino and neutralino masses
\cite{ANc,LNZI}:
\begin{equation}
m_{\chi^\pm_1}\approx m_{\chi^0_2}\approx2m_{\chi^0_1}\ ,\qquad
m_{\chi^0_{3,4}}\approx m_{\chi^\pm_2}\approx~|\mu|\ .
\label{eq:chnt}
\end{equation}
The degree of approximation implied by these mass relations varies somewhat
from model to model. The origin of these mass relations can be traced back
to the relatively large value of $|\mu|$ that follows from the radiative
electroweak symmetry breaking mechanism.

\begin{figure}[p]
\vspace{5in}
\includegraphics{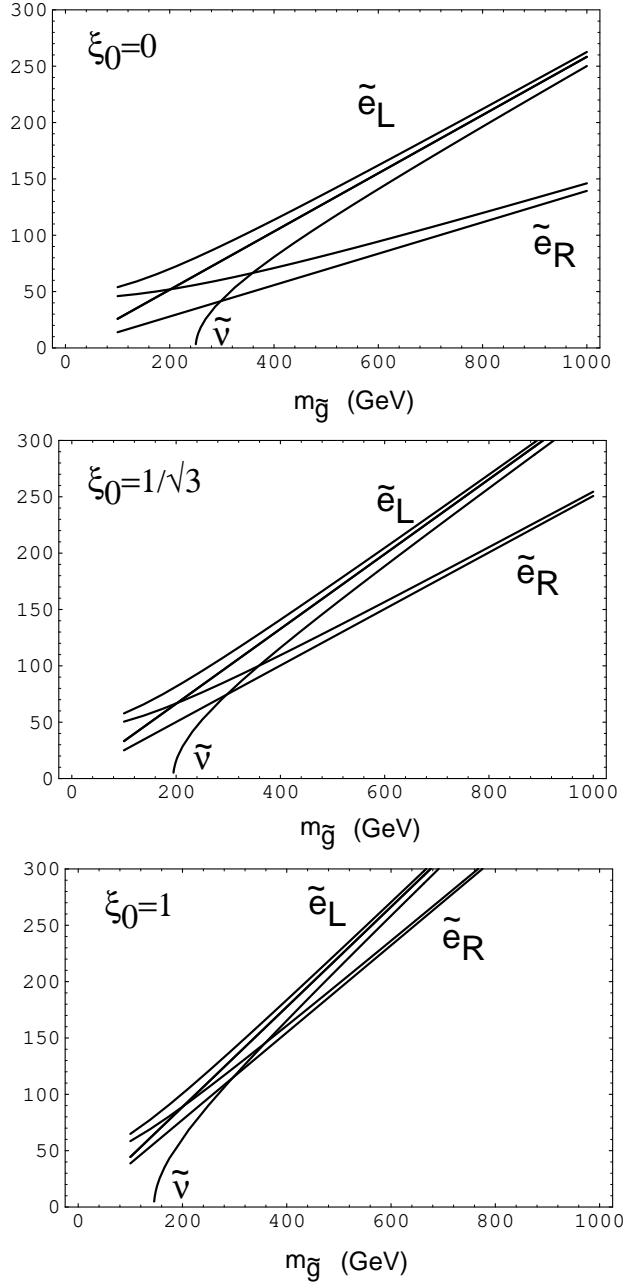}
\vspace{2in}
\caption{Slepton masses as a function of the gluino mass in GUTs, for different
choices of the $\xi_0=m_0/m_{1/2}$ parameter. The straight lines correspond to
$\tan\beta=1$, while the curved lines correspond to $\tan\beta\gg1$.}
\label{fig:sleptonsGUT}
\end{figure}

\begin{figure}[p]
\vspace{5in}
\includegraphics{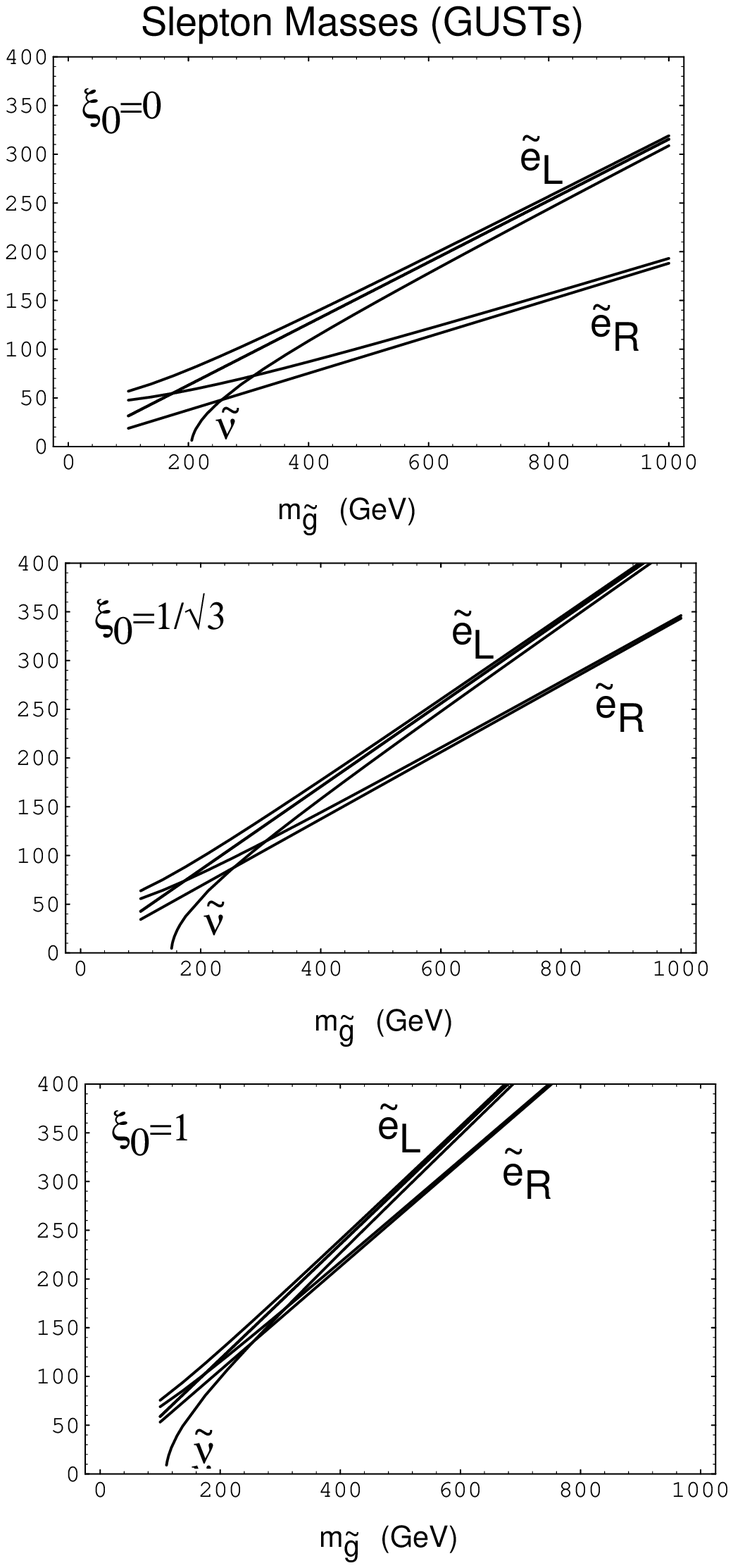}
\vspace{2in}
\caption{Slepton masses as a function of the gluino mass in GUSTs, for
different choices of the $\xi_0=m_0/m_{1/2}$ parameter. The straight lines
correspond to $\tan\beta=1$, while the curved lines correspond to
$\tan\beta\gg1$.}
\label{fig:sleptonsGUST}
\end{figure}

\begin{figure}[p]
\vspace{5in}
\includegraphics{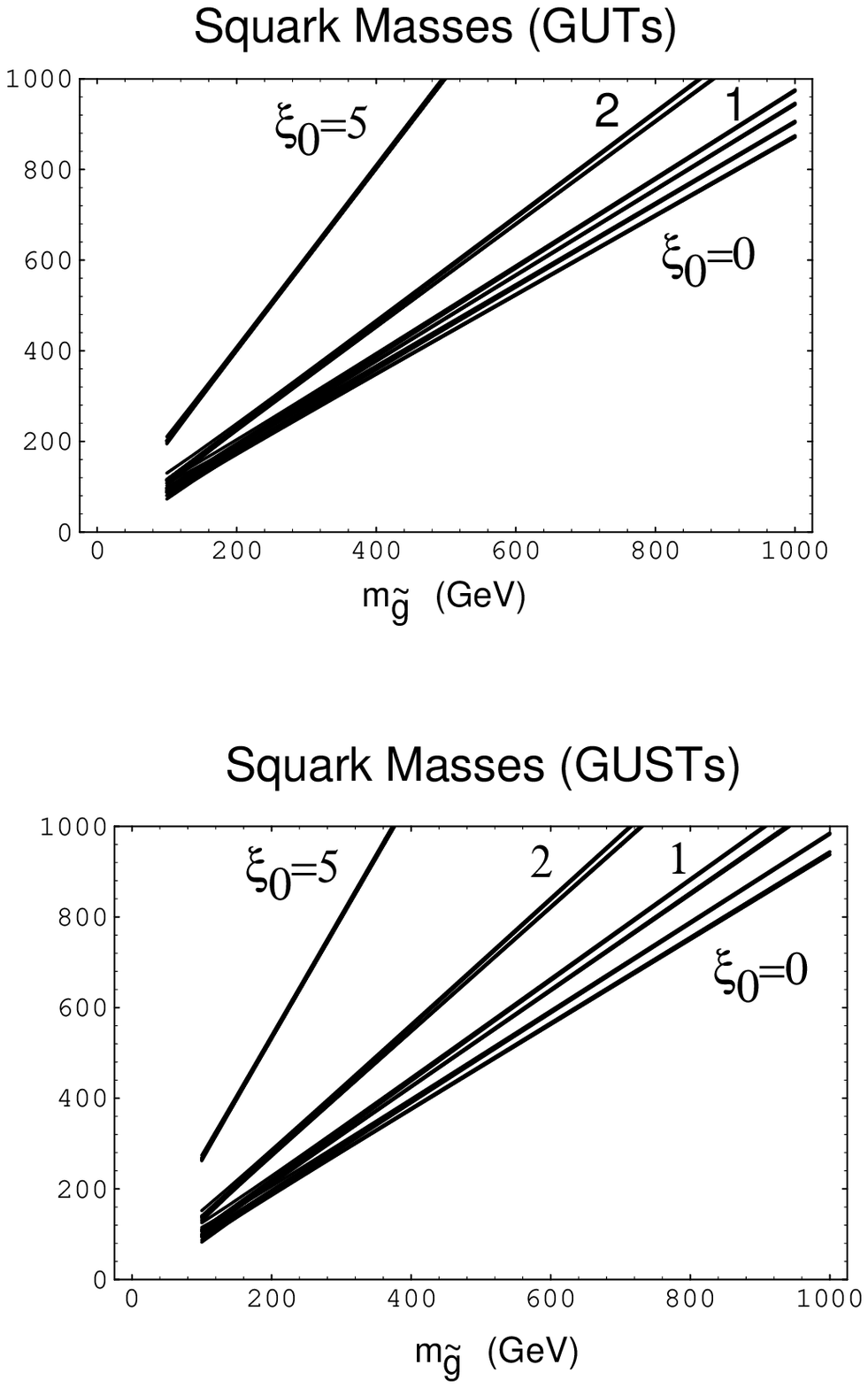}
\vspace{1in}
\caption{Squark masses as a function of the gluino mass in GUTs and GUSTs, for
different choices of the $\xi_0=m_0/m_{1/2}$ parameter. The choices of
$\tan\beta$ ($1,\gg1$), for each value of $\xi_0$, span the whole allowed
range. Note that for $\xi_0\sim1$, $m_{\tilde q}\approx m_{\tilde g}$.}
\label{fig:squarks}
\end{figure}

\subsection{Typical spectra}
To be concrete, in Fig.~\ref{fig:sleptonsGUT} we show the masses of the first-
and second-generation sleptons ($m_{\tilde e_R}=m_{\tilde\mu_R}$, $m_{\tilde
e_L}=m_{\tilde\mu_L}$,$m_{\tilde\nu_e}=m_{\tilde\nu_\mu}$) as a function of the
gluino mass for three choices of $\xi_0=0,{1\over\sqrt{3}},1$; and for
$\tan\beta=1$ (straight lines) and $\tan\beta\gg1$ (curved lines), as
calculated from Eq.~(\ref{eq:masses}) using the numerical coefficients in
Table~\ref{Table3} \cite{LNZI,dilaton}. This exercise is repeated in
Fig.~\ref{fig:sleptonsGUST} for the string-inspired model which unifies at the
string scale. Perhaps the most interesting feature of these figures is the
implied lower bound on the gluino mass from the presently known lower bounds on
the slepton masses from LEP~1 data ($m_{\tilde\ell}\gsim45\GeV$). In particular
the sneutrino mass is quite restrictive. These indirect lower bounds show that
discovery of the gluino at the Tevatron could not have occurred so far, as the
experimental sensitivity has just recently reached the $200\GeV$
range.\footnote{It should be noted that a light gluino window ($m_{\tilde
g}\sim {\rm few}\GeV$) appears to still be allowed experimentally
\cite{LightGluino}, although it may be theoretically disfavored \cite{LNW}.} In
Fig.~\ref{fig:squarks} we present the analogous plots for the squark masses.
Note that the masses of these first- and second-generation squarks are nearly
degenerate (as indicated above) with the main dependence embodied in the
parameter $\xi_0$. Note also that for $\xi_0\sim1$, we obtain $m_{\tilde
q}\approx m_{\tilde g}$, which is the region in the $(m_{\tilde q},m_{\tilde
g})$ plane of greatest experimental sensitivity. These figures also show that
unless $\xi_0\gg1$, the slepton masses are expected to be much lighter than the
squark masses. From Eqs.~(\ref{eq:gauginos}) and (\ref{eq:chnt}) one can show
that
\begin{equation}
m_{\chi^\pm_1}\approx M_2\approx0.3 m_{\tilde g}\ ,
\label{eq:chg}
\end{equation}
and therefore the weakly-interacting charginos and neutralinos are much lighter
than the strongly interacting squarks and gluino. Moreover, if one imposes
an upper limit on the squark and gluino masses of $1\TeV$, the corresponding
upper limit on the lighter chargino and neutralinos is under $300\GeV$.

We should also comment on the Higgs-boson mass spectrum. Because of the
constraints from radiative electroweak symmetry breaking \cite{LNPWZh}, which
effectively link the sparticle and Higgs sectors of the theory, as the
supersymmetry-breaking scale is raised, the lightest Higgs boson mass ($m_h$)
approaches its asymptotic value, as determined by the one-loop expression in
Eq.~(\ref{eq:mhloop}). For $m_t\lsim180\GeV$ one obtains $m_h\lsim130\GeV$.
The remaining Higgs bosons ($A,H,H^\pm$) acquire a mass close to $|\mu|$, and
decouple from the fermions and gauge bosons (their couplings are suppressed).
Moreover, the couplings of the lightest Higgs boson ($h$) to fermions and gauge
bosons approach those of the Standard Model Higgs boson in this limit.
Therefore, it becomes rather difficult to distinguish between these flavors of
Higgs bosons, except for new supersymmetric decays of $h$ into the lightest
supersymmetric particle ($h\to\chi^0_1\chi^0_1$), which will erode the
preferred $h\to b\bar b$ mode when kinematically allowed.

\section{Experimental Prospects}
\label{sec:prospects}
The most basic experimental predictions of supersymmetric models, \ie, the type
of particles to be found and their coupling strengths, are to a great extent
fixed simply by the presence of supersymmetry. However, a quantification of
supersymmetry breaking is essential to determine the masses of the
superparticles, and therefore their discovery windows at experimental
facilities. Unless one is dealing with the straight MSSM, where all
superparticle masses are to be taken as independent parameters (something that
is usually not done in practice anyway), the various levels of theoretical
input that we have discussed above lead to a vast number of {\em correlated}
experimental predictions. The popular models based on universal
soft-supersymmetry-breaking can be described in terms of only four parameters.
More detailed models require even less parameters, and in principle no
parameters.

One can search for supersymmetry directly at collider experiments such as
Fermilab's proton-antiproton Tevatron collider and its proposed upgrades,
CERN's electron-positron LEP collider, CERN's proposed Large-Hadron-Collider
(LHC), the proposed Next-Linear-Collider (NLC), the proposed
First-Muon-Collider (FMC), etc. In the time frame of 1996--2006 one expects
to see the completion of the Tevatron program (Runs II and III), the completion
of the LEP program, the start of the LHC program, and a definite timetable for
the NLC (and perhaps even the FMC). One can also search for supersymmetry
through indirect effects which may affect the expected predictions of certain
Standard Model processes. Such precision measurents will be carried out at
CLEO ($b\to s\gamma$), Brookhaven (g-2), the Tevatron and LHC (rare top-quark
decays), B-factories (CP violation), proton decay experiments, KTeV (rare kaon
decays), cryogenic dark matter detectors (direct detection of dark matter) and
neutrino telescopes (indirect detection of dark matter), neutrino oscillation
experiments (neutrino masses and mixings), etc.

\subsection{Direct Searches}
\label{sec:DirectSearches}
\subsubsection{Hadron Colliders}
\begin{figure}[t]
\vspace{4in}
\includegraphics{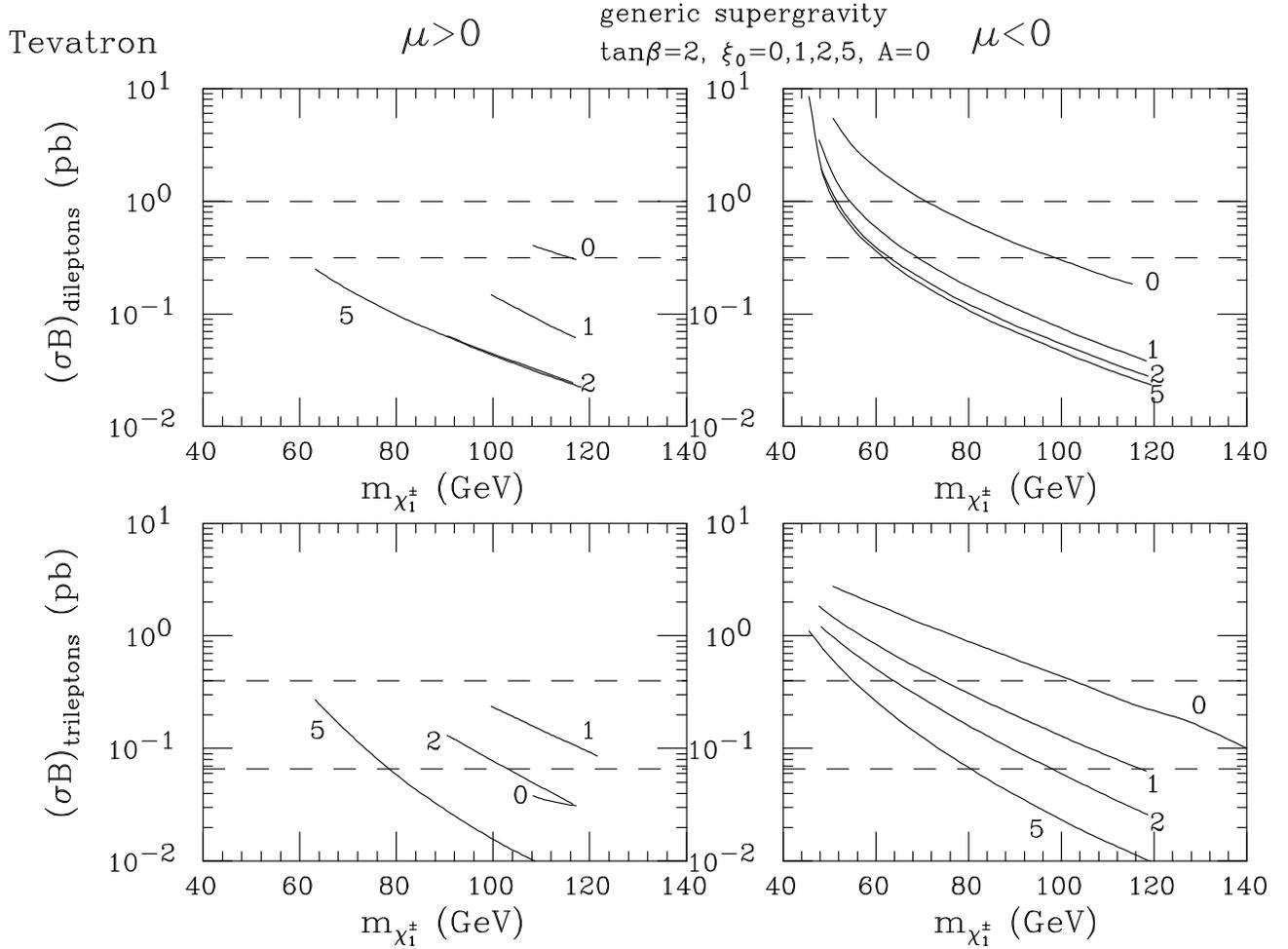}
\vspace{2.6cm}
\caption{The dilepton and trilepton rates at the Tevatron versus the chargino
mass in a generic unified supergravity model with $\tan\beta=2$,
$\xi_0=0,1,2,5$ (as indicated), and $A=0$. The upper (lower) dashed lines
represent estimated reaches with $100\,{\rm pb}^{-1}$ ($1\,{\rm fb}^{-1}$) of
data.}
\label{fig:di-trilep}
\end{figure}

Supersymmetric particles have been searched for since 1988 in $p\bar p$
collisions at $\sqrt{s}=1.8\TeV$ by the CDF and D0 Collaborations
at the Tevatron (Run I), with a total integrated luminosity $\sim100\ipb$
at the end of 1995. Early on the preferred signature was that of jets plus
missing energy, as predicted to occur in the production and decay of the
strongly-interacting gluino and squarks ($p\bar p\to \tilde q\tilde q,\tilde
g\tilde g,\tilde q\tilde g$). Such signature has not been seen, and lower
bounds of $m_{\tilde q,\tilde g}\gsim 175\GeV$ and $m_{\tilde q}\approx
m_{\tilde g}\gsim230\GeV$ have been set \cite{TeVsqg}. More recently it has
been realized that since the practical reach into squark and gluino masses has
been nearly reached, one should also consider the production of
weakly-interacting superparticles (charginos and neutralinos: $p\bar
p\to\chi^\pm_1\chi^0_2+X$), which have smaller cross sections, but that are
typically expected to be much lighter than squarks and gluino. This endeavor
has benefited greatly from the existence of a nearly background-free decay of
the chargino and neutralino into three charged leptons
\cite{Oldtrileptons,Newtrileptons,KLMW,di-tri}. Chargino pair-production into
dileptons has also been considered recently \cite{di-tri}. Preliminary results
have since appeared \cite{D03l}. With the full data set for Run I, it is
expected that the Tevatron could probe chargino masses as high as
$\sim100\GeV$, in some regions of parameter space. Trilepton and dilepton rates
as a function of the chargino mass in a generic supergravity model
\cite{di-tri} are displayed in Fig.~\ref{fig:di-trilep}. It should be noted
that there are regions of parameter space where the trilepton rate is
negligible, due to the presence of ``spoiler" modes that overwhelm the
trilepton signal \cite{Newtrileptons}. The Tevatron should also be able to set
new lower bounds on light top squarks \cite{BST}; first experimental results
have recently appeared \cite{D0stop}.

The Tevatron is expected to be upgraded significantly  with the commissioning
of the Main Injector (1999), which will allow accumulation of integrated
luminosities of a few $\ifb$. Supersymmetry searches will continue in this
upgraded machine, with modest gains expected in the squark-gluino sector, but
great improvements expected in the chargino-neutralino sector (see
Fig.~\ref{fig:di-trilep}). Yet further into the future (2002) a high-luminosity
Tevatron may be in operation (TeV33) \cite{TeV33}, entailing further
exploration of the parameter space, mostly in the chargino-neutralino sector,
and perhaps also in the Higgs sector \cite{HiggsTeV}. (A doubling of the
Tevatron energy (the DiTevatron) has also been considered \cite{KLMW}.) Around
2004 one expects the commissioning of the LHC, where the searches for gluino
and squarks in 14 TeV $pp$ collisions will again take center stage. The LHC
should reach easily into the TeV mass region for these particles
\cite{BaerLHC}. Other supersymmetry searches at the LHC are more uncertain,
given the extremely high collision rate and multiple-interaction environment.
Detection of light Higgs bosons may also pose a problem \cite{HiggsLHC}. In any
event, the LHC is expected to be the definitive experiment for low-energy
supersymmetry: either it will be observed there (or before) or it will be
rendered rather unappealing as an extension of the Standard Model.

\subsubsection{Lepton Colliders}
Soon after its commissioning in 1989, LEP~1 data on the total width of the
$Z$ boson showed that new particles with unsuppressed couplings to the $Z$
boson had to had masses larger than $\sim{1\over2}M_Z$ \cite{LEP1Limits}. This
limit applies to most supersymmetric particles. Exceptions may include the
lightest neutralinos \cite{LEP1chi0} and the lightest top-squark, whose
coupling to the $Z$ may be suppressed in some regions of parameter space. LEP~1
also searched for the Standard Model Higgs boson via the process: $e^+ e^-\to
Z\to Z^*H$ and obtained the limit $M_H\gsim65\GeV$ \cite{LEP1Higgs}. This limit
tends to apply to the lightest Higgs boson ($h$) in supergravity models,
especially when the sparticle spectrum is in the few hundred GeV range
\cite{LNPWZh}. Otherwise, the limit in the MSSM is weaker ($m_h\gsim40\GeV$,
$m_A\gsim20\GeV$). The LEP program is undergoing an energy upgrade, with the
near-term goal of reaching the threshold for production of $WW$ pairs in mid
1996, and an eventual goal of reaching a center-of-mass energy of 192 GeV by
1998. An intermediate-energy step (``LEP~1.5") with $\sqrt{s}=130-136\GeV$,
accumulated a $\sim6\ipb$ of data in November of 1995, and was able to increase
the lower bound on the chargino mass up to 64 GeV (in most regions of parameter
space) \cite{LEP1.5}. With the full-energy upgrade ($\sqrt{s}\sim200\GeV$), the
LEP~2 program should be able to push the lower limits on sparticle masses to
near the kinematical limit \cite{LEPsusy}. The Higgs boson will be searched for
via the process $e^+e^-\to Z^*\to Zh$ \cite{EGN}, with an expected reach
strongly dependent on $\sqrt{s}$ and the accumulated integrated luminosity
(\eg, $m_h\sim\sqrt{s}-100\GeV$ for $500\ipb$ of data) \cite{Sopczak}. After
sparticles are discovered and their masses approximately determined, the NLC
should be able to follow a program of sparticle spectroscopy that will
determine the spectrum of supersymmetric particles rather precisely
\cite{FHKN,Peskin}.

\subsection{Indirect Searches}
\label{sec:IndirectSearches}
\subsubsection{Precision measurements}
Precision measurements at the $Z$ pole have shown that the Standard Model
works rather well; deviations from it should be naturally suppressed, as is the
case of supersymmetry. Nonetheless, LEP~1 has left us with two puzzles: the
measured value of $\alpha_s(M_Z)$ appears to be 10\% higher \cite{bethke} than
that inferred from low-energy measurements \cite{Shifman}, and the
$R_b=\Gamma(Z\to b\bar b)/\Gamma(Z\to{\rm hadrons})$ observable appears
inconsistent with the Standard Model prediction at the 3$\sigma$ level
\cite{LEP}. These discrepancies may disappear with a better understanding of
the experimental procedures, or they may signal the presence of new physics
beyond the Standard Model. Supersymmetry offers some hope to explain both of
them, but only if superparticles are very light \cite{KSW,Rb}. LEP~2 searches
will determine whether these particular regions of parameter space remain
viable or not. In fact, LEP~1.5 searches imply that the $R_b$ anomaly is
unlikely to be due to the presence of light supersymmetric particles
\cite{ELNRb}.

The measurement of the one-loop FCNC decay $B(b\to s\gamma)$ by the CLEO
Collaboration \cite{CLEO} agrees well with the Standard Model prediction and
has constrained supersymmetric models in significant ways \cite{bsgamma},
as exemplified in Fig.~\ref{fig:bsg}. In particular, large values of
$\tan\beta$ appear now disfavored. Future indirect test of supersymmetry
include refinements of the $B(b\to s\gamma)$ measurement (although the
theoretical prediction may not have the corresponding precision \cite{Buras}),
and the new Brookhaven $(g-2)_\mu$ experiment \cite{Roberts}, which is expected
to constrain supersymmetric models in significant ways \cite{g-2susy}
(especially for large values of $\tan\beta$) once it starts to take data in
1996.

\begin{figure}[t]
\vspace{4in}
\includegraphics{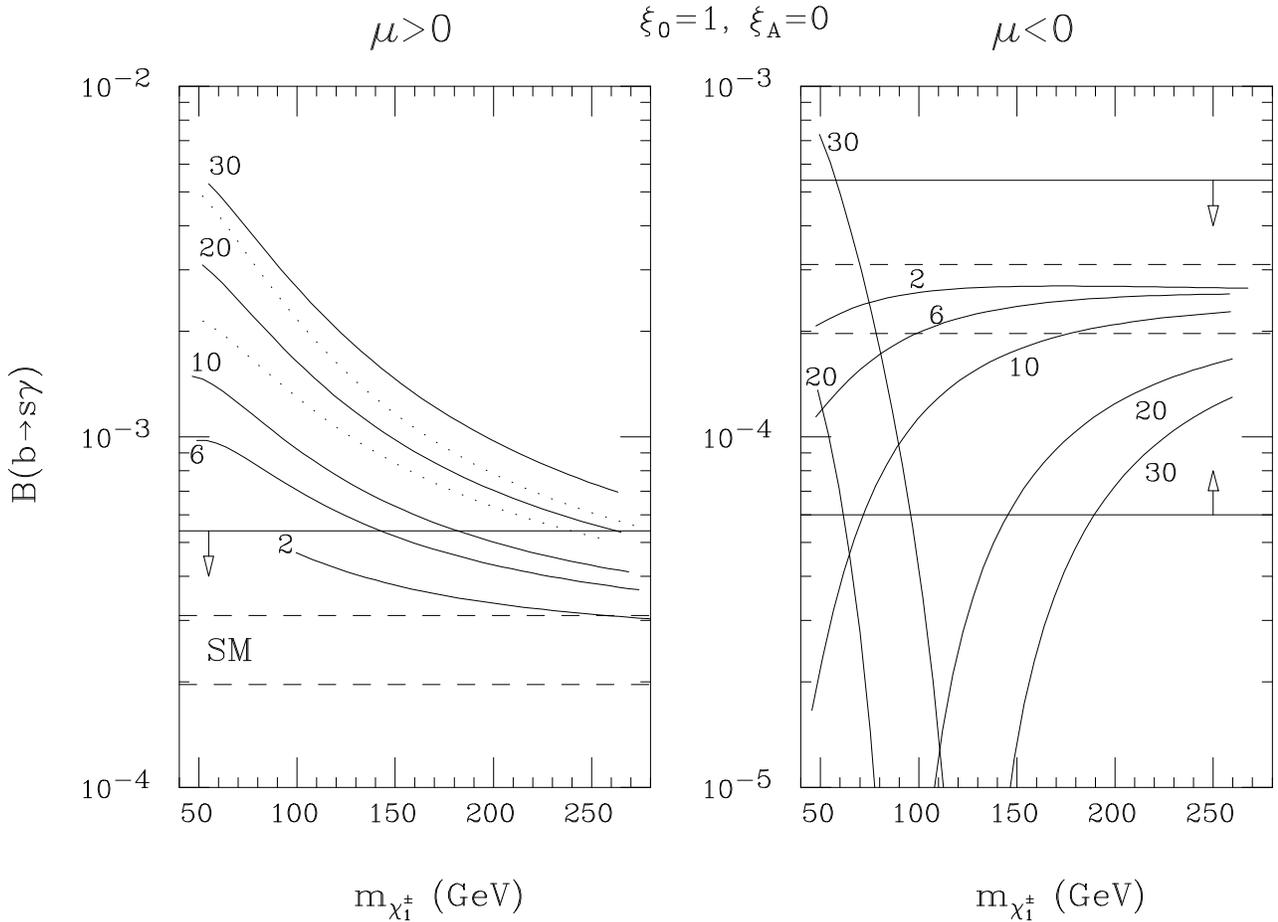}
\vspace{2cm}
\caption{The calculated ``central" values of $B(b\to s\gamma)$ in a
generic supergravity model for $\xi_0=1$ and $\xi_A=0$, for representative
values of $\tan\beta$. The dotted curve above (below) the $\tan\beta=20$
curve for $\mu>0$ corresponds to $\xi_A=-1\,(+1)$. The arrows point into the
experimentally allowed region. The dashed lines delimit the Standard Model
range.}
\label{fig:bsg}
\end{figure}

\subsubsection{CP, CPT, and LFV}
Tests of CP violation are particularly relevant to supersymmetric models,
as these models typically include new phases beyond that in the CKM matrix
\cite{DGH}. The most stringent test has been that of $K$-$\bar K$ mixing
($\Delta M$ and $\epsilon_K$), which demands that squarks of the same electric
charge (but different flavor) should be nearly degenerate in mass
\cite{KKsusy}. This requirement can be naturally accommodated in supergravity
models with universal supersymmetry breaking masses at the unification scale,
and provides an important restriction on novel string-based scenarios where
universality is typically absent \cite{Ibanez,FCNCstrings}. Rare kaon decays
(CP violating or not) are also sensitive probes of supersymmetry
\cite{RareKsusy}. Lepton-flavor-violating (LFV) processes are expected in
unified models, as this Standard Model symmetry is typically violated in GUT
models \cite{LFV}. String models generally predict the violation of CPT
symmetry \cite{CPTx} because of their inherent lack of locality at very small
distances. The $K$-$\bar K$ system \cite{ELMN} (and perhaps also the $D$ and
$B$ systems \cite{KP2}) appears to be particularly sensitive. The above kind of
processes have been studied at Brookhaven, Fermilab, and CERN, and will
continue in the future including new players such as KTeV \cite{KTeV}, the SLAC
B factory \cite{Bfac}, DAFNE \cite{DAFNE}, etc.

\subsubsection{Proton decay}
Proton decay is an unambiguous test of new physics at the unification scale
\cite{Oldpd}, and has in the past excluded altogether the original SU(5) model
via the $p\to e^+\pi^0$ mode \cite{PDG}. The minimal SU(5) supergravity model
has also been challenged through the inherently supersymmetric $p\to\bar\nu
K^+$ mode \cite{Newpd}. Future tests will commence as soon as SuperKamiokande
goes online in 1996. The  expected sensitivity should provide a definitive test
of SU(5) GUTs (via $p\to\bar\nu K^+$) \cite{Newpd}. Moreover, the traditional
$p\to e^+\pi^0$ mode may constrain flipped SU(5), should the strong coupling
$\alpha_S(M_Z)$ settle near 0.110 \cite{lowering}.

\subsubsection{Neutrino oscillations}
Evidence for non-zero neutrino masses may constitute the first deviation from
the Standard Model. Unified models have a natural mechanism for producing
such small masses (the see-saw mechanism), and even provide detailed
predictions for the pattern of neutrino masses and mixings \cite{Langacker}.
There is a plethora of presently (Homestake, Kamiokande, SAGE, GALLEX)
or soon-to-be operating (SNO, Superkamiokande, Borexino) solar neutrino
detectors, atmospheric neutrino detectors (Kamiokande, IMB, Soudan II),
and neutrino-oscillation experiments (LSND, CHORUS, NOMAD, E803), which are
providing data that can be interpreted in this theoretical context. The data
has so far been insufficient to declare the discovery of neutrino masses or
neutrino oscillations, although the solar neutrino data appears to be most
naturally explained by the MSW mechanism \cite{MSW} of matter-enhanced
oscillations in the Sun \cite{HL} with neutrino mass and mixing parameters
that are readily obtained in GUTs. Atmospheric neutrino data and the Los Alamos
(LSND) neutrino oscillation experiment \cite{LSND} are more difficult to
interpret in this theoretical context  \cite{Mohapatra}. The theoretical
situation should become more clear once enough data has been accumulated on
each category of experiments to be able to assess their statistical
significance.

\subsubsection{Dark matter detectors}
Supersymmetric models that respect $R$ parity (a popular and well motivated
assumption) possess a natural candidate for cold dark matter, the lightest
supersymmetric particle, which must be neutral and colorless \cite{EHNOS} and
is usually assumed to be the lightest neutralino. (The sneutrino may happen to
be the LSP, but this dark matter candidate is severely constrained.) Such
stable particle would populate the galactic halo and could be detected directly
or indirectly \cite{Griest}. Direct dectection experiments \cite{lspd}
rely on energy deposited in cryogenic detectors after a direct
nucleon-neutralino collision in the detector. Indirect detection relies on
the capture of neutralinos by the Sun or Earth \cite{NTold}, and their
subsequent annihilation into (among other things) energetic neutrinos, which
can be detected in underground or underwater facilities (``neutrino
telescopes") \cite{NTnew}. Both these detection mechanisms are quite promising
\cite{DMrecent} and a number of facilities of both kinds are currently in
operation or will soon be operating, including the Stanford Germanium detector,
Kamiokande and SuperKamiokande, MACRO, Amanda, Nestor, DUMAND, etc.

\section{Conclusions}
\label{sec:conclusions}
The Standard Model of elementary particle physics has been subjected to
intense experimental scrutiny over the last twenty years. With the advent of
the Tevatron and LEP colliders, these tests have reached unprecedented
precision. Yet, in nearly all instances observations agree very well with
theoretical expectations. This state-of-affairs indicates that whatever new
physics may lie beyond the Standard Model, it will not be intertwined with
it in any significant way. This is a very useful fact, as Standard Model
processes will constitute the bulk of the events in searches for new physics
at higher energies. However, because of the ``purity" of the Standard Model,
these background processes will be reliably calculated, and the signals for
new physics will be more easily extractable. Such searches are underway and
will reach new sensitivity levels in the near future, most notably with the
LEP~2 energy upgrade, the Main Injector Tevatron upgrade, and in the long-term
the Large Hadron Collider; and indirectly via searches for rare processes
such as proton decay at SuperKamiokande and the anomalous magnetic moment of
the muon at Brookhaven.

A very well motivated possibility for the type of new physics that one may
encounter has been the subject of this review: supersymmetry. As we have
discussed, supersymmetry is an underlying theme in the march towards ever
increasing energy scales. In fact, it appears to be the only road that allows
us to see the light at the end of the tunnel. There will always be
alternatives, but they must all contend with the gauge hierarchy problem, and
supersymmetry is the only known way of tackling it without giving up
calculability. Interestingly enough, despite all of the impetus with which
these ideas have been pursued, direct evidence for the existence of
supersymmetry is yet to be found. However, I believe that supersymmetry will
not run away from us for much longer. In fact, our best models today indicate
that the success of the Standard Model effectively pushes supersymmetry up
to higher mass scales which are just now beginning to be explored
experimentally.

On the other hand, supersymmetry has received a great deal of indirect
evidence over the last several years. Most strikingly was the convergence of
the precisely measured Standard Model gauge couplings at very high energies.
This fact led to a revival of these ideas, including intense theoretical
scrutiny of gauge and Yukawa coupling unification in unified supergravity
models. It has also been claimed that the two known experimental ``anomalies"
in the Standard Model -- the discrepancy between measurements of the strong
coupling at LEP and their counterparts at low energies, and the discrepancy
between the measurement of and the Standard Model prediction for $R_b$ -- may
reflect the presence of new physics and in particular the existence of light
supersymmetric particles.

Another bit of indirect evidence comes from the discovery of the top quark with
a mass $m_t\sim200\GeV$. There is only one known theory where quark masses can
be calculated, namely string theory. In string models the Yukawa coupling
that gives rise to the top-quark mass is naturally of the size needed to
yield such ``large" quark masses. Moreover, string models also explain
the lightness of the remaining quark masses, as other Yukawa couplings tend
to be suppressed by stringy selection rules. Thus, a large top-quark mass can
only be understood in the context of string theory. Furthermore, a large
top-quark mass can only be predicted in the presence of supersymmetry, as one
needs to connect stringy predictions at the Planck scale with top-quark masses
measured in the laboratory. String theory itself is undergoing a second
revolution (the first one occurred in 1984 with the establishment of string
theory itself) with many new relations being found among previously thought
distinct types of string. Strings also appear to be intimately connected with
higher-dimensional ``membranes", leading to the conjecture of a universal
M-theory underlying all the different phases of strings. Hopefully these
developments will shed light onto the unresolved problems of supersymmetry
breaking and the determination of the vacuum. In fact, further progress in
supersymmetry model building should come from string models, as these provide
us with the ability to calculate.

Supersymmetry will continue to have an ever expanding role in the physics of
the very early universe, and its present-day manifestations. The observed
minute anisotropies in the cosmic microwave background radiation show an
imprint left early on, which points towards the idea of inflation where
GUT- and Planck-scale physics play a major role. It has also
become clear that most of the matter in the universe is invisible. Moreover,
conventional astrophysical explanations (red dwarfs, brown dwarfs, etc.) have
been found to constitute only a small fraction of this ``dark" matter.
Supersymmetry comes in again by providing a natural candidate for such dark
matter, which may in fact be detectable in the laboratory.\\

We should soon know whether supersymmetry is within our reach or not.

\section*{Acknowledgements}
\noindent This work has been supported in part by DOE grant
DE-FG05-93-ER-40717.

\end{document}